\documentclass[a4paper,11pt]{article}
\pdfoutput=1 
\usepackage{jcappub}
\usepackage{graphicx}
\pdfoutput=1 
\usepackage{color}
\usepackage{makecell}
\usepackage{afterpage}
\usepackage[dvipsnames]{xcolor}
\usepackage{multirow}
\usepackage{cleveref}
\usepackage[normalem]{ulem}		
\usepackage{xspace}
\usepackage{subcaption}
\usepackage{booktabs}

\usepackage{centernot}
\usepackage{mathtools}
\usepackage{stmaryrd}

\definecolor{cadmiumgreen}{rgb}{0.0, 0.42, 0.24}

\newcommand{\fulleqref}[1]{equation~\eqref{#1}}
\newcommand{\fullsecref}[1]{section~\ref{#1}}
\newcommand{\fullmanyeqref}[1]{equations~\eqref{#1}}

\newcommand{\tquote}[1]{``#1''}

\newcommand{\dnew}{\par\vspace*{0.5\baselineskip}\mbox{}}
\newcommand{\dnewnoindent}{\par\vspace*{0.5\baselineskip}\noindent\mbox{}}

\newcolumntype{L}[1]{>{\raggedright\arraybackslash}p{#1}}
\newcolumntype{C}[1]{>{\centering\arraybackslash}p{#1}}

\newcommand{\class}{{\sc class}\xspace}

\title{CosmicNet I: Physics-driven implementation of neural networks within Boltzmann-Einstein solvers}

\author[1]{Jasper Albers,}
\author[1]{Christian Fidler,}
\author[1]{Julien Lesgourgues,}
\author[1]{Nils Sch\"oneberg,}
\author[1]{and Jesus Torrado}

\affiliation[1]{Institute for Theoretical Particle Physics and Cosmology (TTK), \\ RWTH Aachen University, D-52056 Aachen, Germany.}

\emailAdd{albers@physik.rwth-aachen.de}
\emailAdd{fidler@physik.rwth-aachen.de}
\emailAdd{lesgourg@physik.rwth-aachen.de}
\emailAdd{schoeneberg@physik.rwth-aachen.de}
\emailAdd{torrado@physik.rwth-aachen.de}

\abstract{Einstein-Boltzmann Solvers (EBSs) are run on a massive scale by the cosmology community when fitting cosmological models to data. We present a new concept for speeding up such codes with neural networks. The originality of our approach stems from not substituting the whole EBS by a machine learning algorithm, but only its most problematic and least parallelizable step: the integration of perturbation equations over time. This approach offers two significant advantages: the task depends only on a subset of cosmological parameters, and it is blind to the characteristics of the experiment for which the output must be computed (for instance, redshift bins). These allow us to construct a fast and highly re-usable network. In this proof-of-concept paper, we focus on the prediction of CMB source functions, and design our networks according to physical considerations and analytical approximations. This allows us to reduce the depth and training time of the networks compared to a brute-force approach. Indeed, the calculation of the source functions using the networks is fast enough so that it is not a bottleneck in the EBS anymore. Finally, we show that their accuracy is more than sufficient for accurate MCMC parameter inference from Planck data. This paves the way for a new project, CosmicNet, aimed at gradually extending the use and the range of validity of neural networks within EBSs, and saving massive computational time in the context of cosmological parameter extraction.}

\begin{document}

\hfill{\small TTK-19-27}

\maketitle

\section{Introduction}
\label{sec:introduction}

{\it Context.} The continuous increase in size and precision of cosmological data triggers an intense world-wide activity of Bayesian comparison of cosmological models to observations. The computational cost of Bayesian model comparison is dominated by cosmology simulation codes. On the one hand, observations probing the non-linear regime of structure formation must be compared with the results of N-body simulations. Such simulations are so computationally expensive that the N-body community discusses and intensively uses several methods to interpolate between N-body results in model parameter space (like emulators, see \cite{Knabenhans:2018cng}). On the other hand, observations probing mainly linear perturbation theory -- like CMB maps, large-scale galaxy surveys,  large-scale cosmic shear surveys or future large-scale 21cm intensity mapping surveys -- only require Einstein-Boltzmann Solvers (EBSs) like {\sc camb} \cite{Lewis:1999bs} or \class \footnote{\url{http://class-code.net}} \cite{Blas:2011rf}, that simulate the evolution of linear perturbations in Fourier space and compute several related observables. For simple tasks (like the computation of CMB anisotropies in a minimal $\Lambda$CDM cosmology including non-zero neutrino masses) these codes run in less than one minute. This is so small compared to an N-body code execution time that less efforts have been devoted to modern interpolator or emulator methods in this context.
\dnew
However, comparing the usage of EBSs and N-body codes, the small cost of a single EBS run is balanced by several other factors. Indeed, the precision of the data on linear scales requires accurate Bayesian inference runs in which millions of models must be evaluated for each run. Besides, the N-body community is forced by feasibility conditions to focus on at most a few extensions of the minimal $\Lambda$CDM model. Instead, the worldwide community exploring new ideas in theoretical cosmology -- and their compatibility with CMB and very-large-scale structure data -- routinely tests hundreds of combinations of extended cosmological models and data sets every week, resulting in tens (or more probably hundreds) of billions of Boltzmann code executions every year. With such gigantic numbers, any approach allowing to reduce the number of EBS executions or their individual CPU time by a significant factor would result in tremendous savings of computational resources (and associated electricity costs). Dramatic gains in efficiency have already been observed in the past, for instance when Seljak and Zaldarriaga \cite{Seljak:1996is} rendered EBSs much more efficient thanks to the line-of-sight integral method, or when Lewis and Briddle \cite{Lewis:2002ah} boosted cosmological parameter inference with Markov Chain Monte Carlo (MCMC) algorithms.
\dnewnoindent {\it State of the art.} Ideally, one would build an emulator trained on EBS results and able to predict approximate cosmological observables at any point in parameter space without actually simulating the cosmological evolution. The idea has been previously implemented with a five-order polynomial interpolation method in the PICO\footnote{\url{https://github.com/marius311/pypico}} project \cite{Fendt:2006uh,Fendt:2007uu}, and with neural networks (NNs) in the CosmoNet\footnote{\url{www.astro.phy.cam.ac.uk/research/ResearchFacilities/software-for-astrophyiscs/cosmonet}} project \cite{Auld:2006pm,Auld:2007qz}. In a project contemporary to this one, Manrique-Yus \& Sellentin use instead neural networks (NNs) to emulate the calculation of galaxy clustering and cosmic shear observables for the Euclid survey~\cite{Manrique}. These attempts target the generation of the observables directly from the input parameters.
\dnew
This simple approach is potentially the one giving the best final performance, with the drawback that a long training is usually needed. Capturing the dependency of a large output data vector on a small input parameter vector is a very complex problem, that usually needs to be tackled by a very fine sampling of the parameter space in the case of interpolators, or by dense and deep networks in the case of neural networks. Besides,  with such methods, the results cannot be extrapolated to variations of parts of the model (e.g., changing the \tquote{fast parameters} of the primordial spectrum while keeping fixed the \tquote{slow parameters} of the cosmological background and thermal evolution \cite{Lewis:2002ah}) or variations of the observable specifications (e.g., the window function of the different redshift bins of cosmic shear or number count $C_\ell$'s, or the modeling of non-linear corrections).
\dnewnoindent {\it New strategy.} In this paper, we take a different approach: instead of targeting the relationship between the input parameters and the observables directly, we look at how the computation of these observables breaks down into simpler steps. We identify the most indivisible and time-consuming one, and we replace exclusively this step by an NN emulator. Additionally, we analyze the physics involved in order to tailor the architecture of the NN to the physics of the model as much as possible.
\dnew
Emulating a simpler computation (than that of the full observable) allows for more analytic control. We can formulate analytic approximations in particular regimes, which can be passed as input to the NN to optimize the network architecture and choose input/output non-linear transformations. This relieves the NN from a lot of the fitting work. It allows us to build shallower networks that retain the full predicting power, since there is less complexity in the parameter dependencies. This directly translates into shorter training and emulation times.
\dnew
Then, when the user faces a new problem (a new extended cosmological model, some new observables, new precision requirements), their interest is to retrain the network before starting Bayesian model comparison, even if very few MCMC runs are needed. Indeed, we will see that retraining the network without changing its overall architecture only takes about half a day on four cores. This overhead can be very quickly compensated, given that afterwards the execution time of the NN-supplemented EBS is significantly reduced, and that each MCMC run typically requires ${\cal O}(10^4)$ or ${\cal O}(10^5)$ calls to the EBS.
\dnew
Another advantage of replacing only one step inside the EBS by an NN (instead of the full code) is that this step depends only on a subset of the parameters governing the final observables. Thus the same NN can be readily used with extended cosmological models that leave the emulated step unchanged (e.g.~in our case, different ansatz on the primordial fluctuation spectra), or with different specifications for the final observables that only play a role in later steps than the emulated one (like bin window functions or non-linear corrections).
\dnew
Both the increase in training speed and the reusability potential of this approach are the selling point of our strategy. Here we implemented our method in \class{}, but our approach is very generic and could be transposed to other codes.
\dnewnoindent {\it Structure of the paper.}  This proof-of-concept work is the first of a series of papers on this topic. We apply our method to the computation of the power spectrum of the Cosmic Microwave Background (CMB). The step that we emulate with NNs is the computation of the CMB source functions (section \ref{sub:sourcefunc}). We describe how we tailor the architecture of the NN to this particular problem in section \ref{sub:network_architecture}, and how we train it in section \ref{sec:training_strategy}. We discuss the accuracy of our implementation at the level of intermediate functions in section \ref{sub:results_source_functions}, of final observables in section \ref{sub:results_power_spectra}, and of the recovery of posterior distributions for input parameters after a full Bayesian parameter extraction in section \ref{sub:results_parameter_extraction}. Finally, we discuss the present and future potential speedup in section \ref{sub:speedup}. We present our conclusions and sketch future developments in section \ref{sec:discussion}.
\dnewnoindent {\it Notations and conventions.} In this paper we will adopt the conventions of naming the comoving Fourier wavenumber $k$, the conformal time $\tau$ ($\tau_0$ being the conformal time today), and using primes for derivatives with respect to conformal time. The Thomson interaction rate $\kappa' = a\,n_e \sigma_T$ is the derivative of the Thomson optical depth $\kappa$ ($\kappa_\mathrm{reio}$ is optical depth to reionization, rather than $\tau_\mathrm{reio}$), while the visibility function reads $g(\tau)= \kappa' e^{-\kappa}$.
\section{Physics-driven network design}

\subsection{Network target: the source functions} \label{sub:sourcefunc}

In this section we are going to motivate our choice to use machine learning to emulate the calculation of the CMB source functions. We will leave the case of other source functions -- like those needed to compute the matter power spectrum $P(k,z)$, the comic shear $C_\ell$'s or the galaxy number count $C_\ell$'s -- for future work.
\dnew
For better readability of this paper, we are going to write the equations of this section assuming a flat FLRW cosmology with exclusively scalar perturbations and adiabatic initial conditions. However, our method does not rely on these assumptions and can be immediately applied to non-flat cosmologies, Bardeen vector or tensor modes, and mixed initial conditions.

\subsubsection{Equations in Boltzmann codes} \label{sub:equations}

The angular power spectra $C_\ell^{XY}$ of the CMB (where $X,Y\in \{T,E\}$ are the temperature or $E$-polarization indices) are given by an integral over the primordial curvature power spectrum $\mathcal{P}_\mathcal{R}(k)$ and photon transfer functions $\Delta_\ell^{X}(k)$:
\begin{equation}\label{eq:angpow tot}
C_\ell^{XY} = 4\pi \int \frac{dk}{k} \mathcal{P}_\mathcal{R}(k) \Delta_\ell^{X}(k)  \Delta_\ell^{Y}(k)~.
\end{equation}
Following the line-of-sight formalism \cite{Seljak:1996is}, the photon transfer functions are given by the convolution of some CMB source functions $S_{X_i}(k,\tau)$ with radial projection functions $\phi_\ell^i(\chi)$, $\epsilon_\ell(\chi)$ (simply related to spherical Bessel functions in a flat universe) given in~\cite{Lesgourgues:2013bra}:
\begin{equation}\label{eq:angpow delta}
\Delta_\ell^{T}(k) = \int d\tau \sum_{i=0,1,2} S_{T_i}(k,\tau) \phi_\ell^i(k\left[\tau_0 - \tau \right])~,
\quad
\Delta_\ell^{E}(k) = \int d\tau S_P(k,\tau) \epsilon_\ell(k\left[\tau_0 - \tau \right])~.
\end{equation}
The CMB source functions depend on transfer functions, i.e.~on perturbations of a given quantity for a given Fourier wavenumber $k$ normalized to an initial curvature perturbation $\mathcal{R}(\mathbf{k})=1$. Adopting the notations of~\cite{Ma:1994dv,Tram:2013ima,Lesgourgues:2013bra}, these include the photon temperature and polarization transfer functions $F_\ell$ and $G_\ell$, the baryon velocity divergence $\theta_b$ and the metric fluctuations. Since CMB source functions are gauge invariant, we can write them in any gauge without loss of generality. In this work, we choose to write equations in the Newtonian gauge, which is more familiar to most readers. The metric fluctuations are then encoded in the two potentials $\phi$ and $\psi$. There are different ways to split the CMB source functions, related to each other through integration by parts. In \class{} these functions are split as~\cite{Lesgourgues:2013bra}:
\begin{subequations}
	\begin{align}
	S_{T_0} &= g \cdot (F_0 + \phi) + e^{-\kappa} 2 \phi' + \left(g \theta_b/k^2\right)'~, \label{eq:ST0} \\
	S_{T_1} &=  e^{-\kappa} k (\psi - \phi)~, \label{eq:ST1}\\
	S_{T_2} &= g \Pi/2~, \label{eq:ST2}\\
	S_P &= \sqrt{6} \left[g \Pi\right]/2~,
	\end{align}
	\label{eq:all_source_functions}
\end{subequations}
\noindent where $\Pi = (G_0 + G_2 + F_2)/4$. Our method would work equally well with other conventions for the splitting of the source functions. Note the very important fact that $S_P = \sqrt{6} S_{T_2}$\,, which is going to allow the network to predict polarization and temperature from the same set of three source functions.
\dnew
The transfer functions are all functions of $(k,\tau)$. For each wavenumber $k$, they are obtained by solving a system of coupled Ordinary Differential Equations (ODEs). 
The evolution of photon perturbations is given by the Boltzmann hierarchy \cite{Ma:1994dv}:
\begin{eqnarray}\label{eq:photonboltzmann}
\delta_\gamma' &=& - \frac{4}{3}\theta_\gamma
+4\phi' \,,\\
\theta_\gamma' &=& k^2 \left(\frac{1}{4}\delta_\gamma
- \sigma_\gamma \right) + k^2 \psi
+ \kappa' (\theta_b-\theta_\gamma) ~,\nonumber\\
F_{\ell}' &=& \frac{k}{2\ell+1}\left[ \ell
F_{(\ell-1)} - (\ell+1)F_{(\ell+1)}
\right] + \kappa' \left[ - F_{\ell} +  \frac{2 \Pi}{5} \delta_{\ell 2} \right] ~,\qquad \text{for} \quad \ell \geq 2 ~,\nonumber\\
G_{\ell}' &=& \frac{k}{2\ell+1}\left[ \ell
G_{(\ell-1)} - (\ell+1) G_{(\ell+1)}\right]
+ \kappa' \left[ -G_{\ell} +  \frac{2 \Pi}{5}\delta_{\ell 2} +  2 \Pi
\delta_{\ell 0} \right]~,\nonumber
\end{eqnarray}
where $\delta_\gamma = 4 F_0$, $\theta_\gamma = 3 k F_1/4$, and $\sigma_\gamma = F_2/2$\,.  The photon hierarchy couples to the perturbation equations of all other species due to Thomson scattering and gravitational interactions. The full system of differential equations can be solved as in \cite{Blas:2011rf}. The results are combined to calculate the CMB source functions~(\ref{eq:all_source_functions}), which can be integrated via \fullmanyeqref{eq:angpow delta} and \eqref{eq:angpow tot} to give the CMB angular power spectra of temperature and polarization anisotropies.
\begin{figure}[tp]
\includegraphics[width=0.49\textwidth]{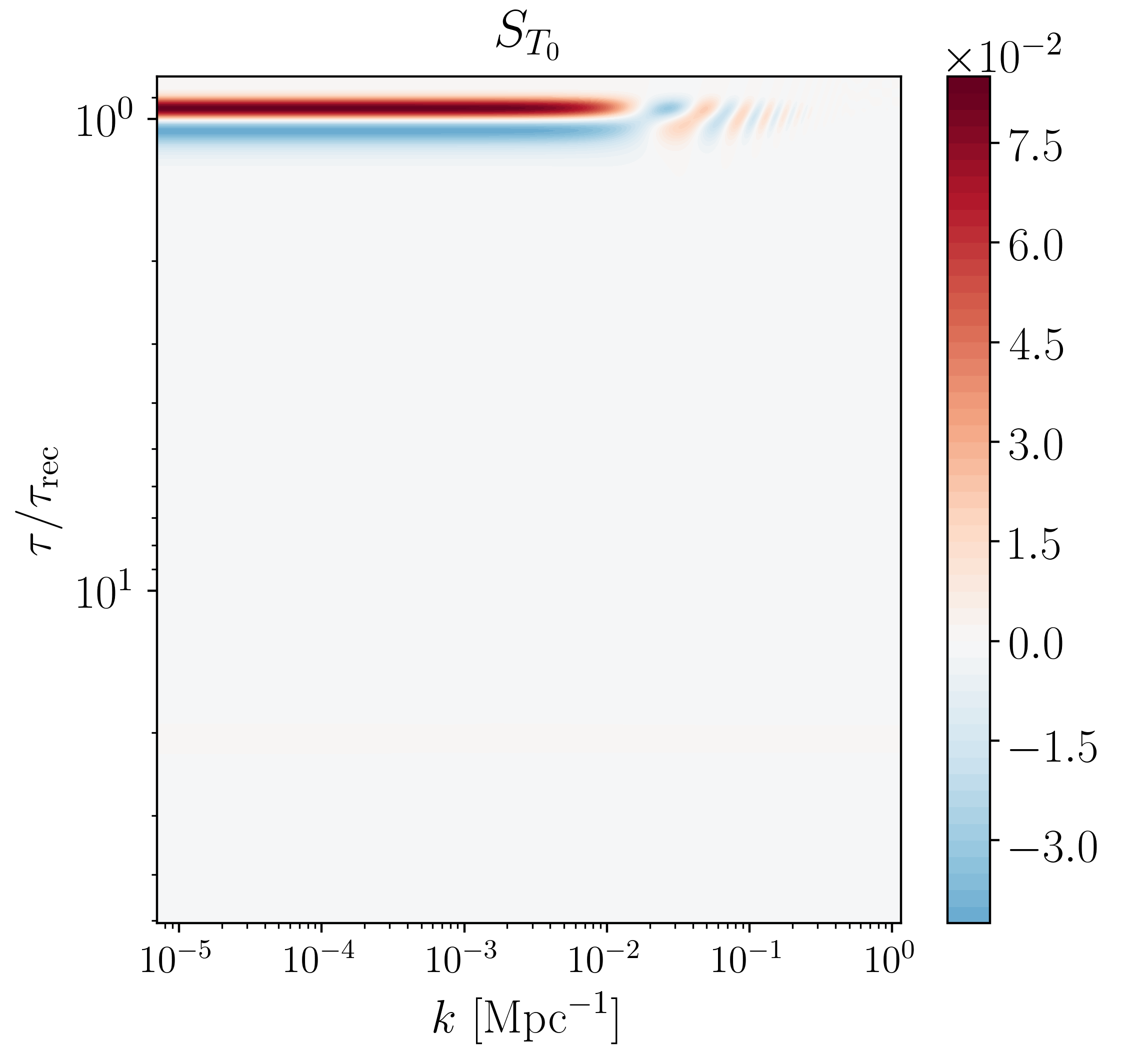}
\includegraphics[width=0.49\textwidth]{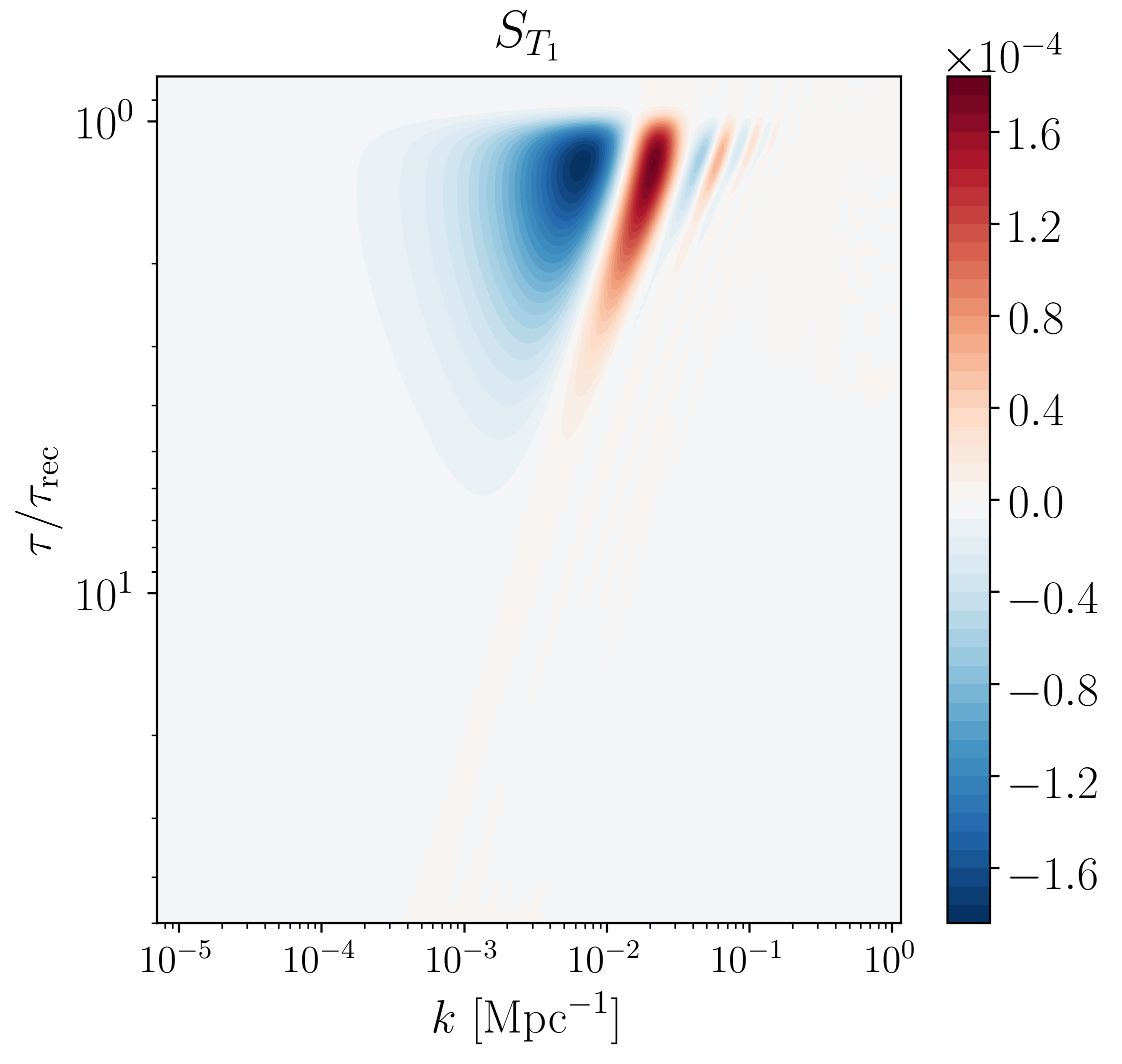}\\
\includegraphics[width=0.49\textwidth]{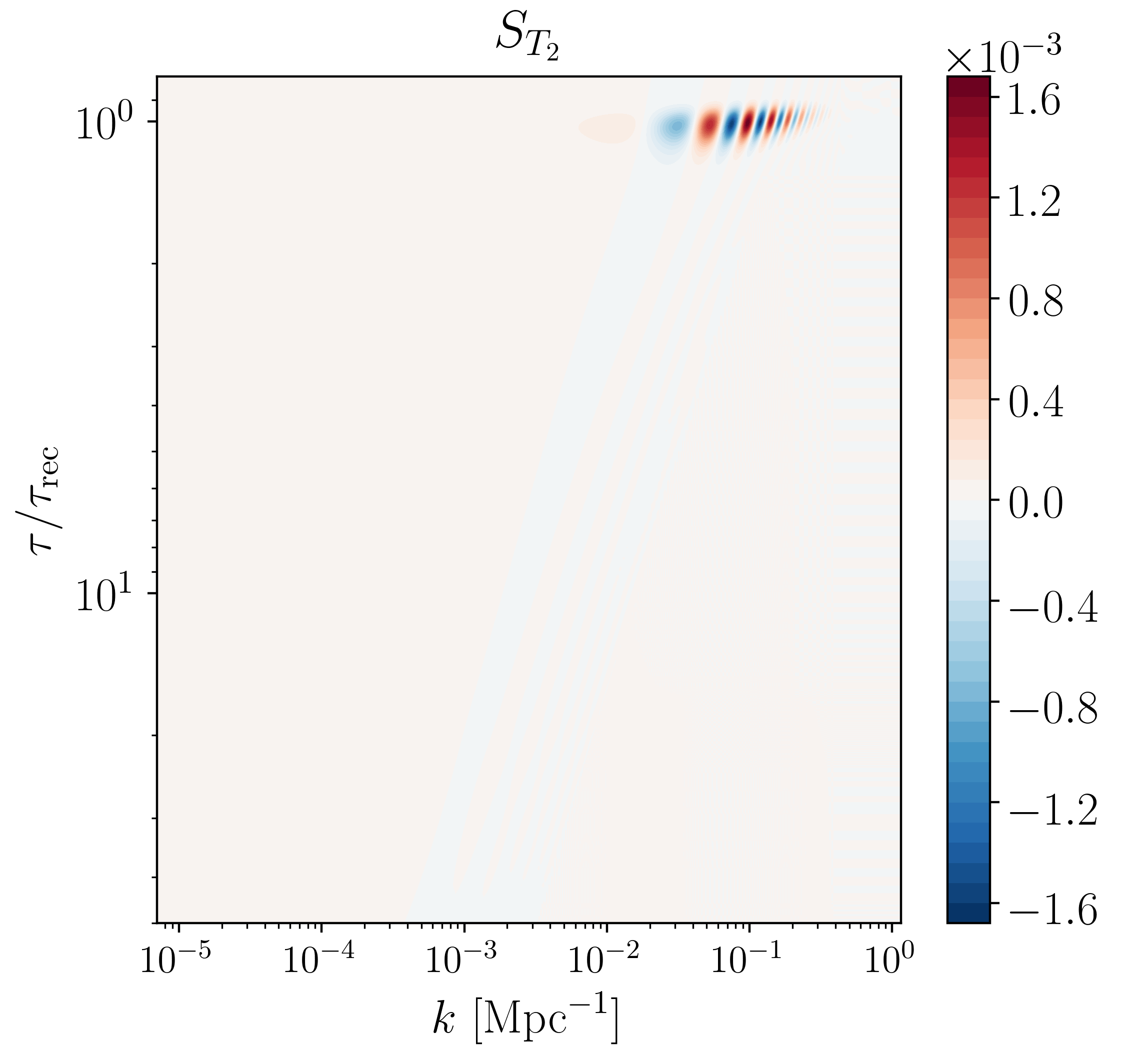}
\includegraphics[width=0.49\textwidth]{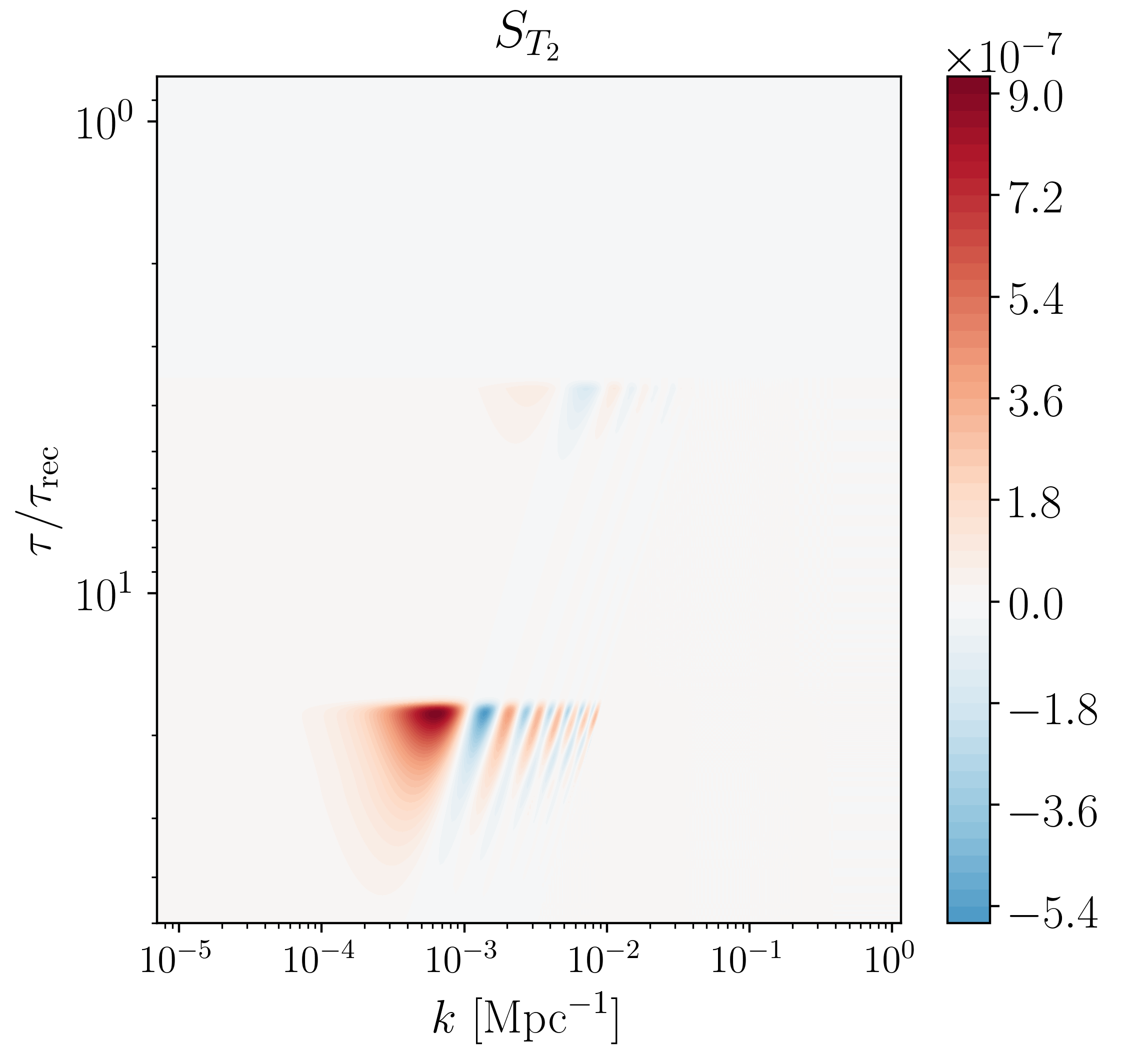}
\caption{CMB source functions $S_{T_1}$, $S_{T_2}$ $S_{T_3}$ as a function of wavenumber $k$ and conformal time $\tau$ (normalized to the recombination time $\tau_\mathrm{rec}$). In the last panel we have enhanced the late-time structure in $S_{T_2}$ by cutting the function below $\tau/\tau_\mathrm{rec} < 3.5$ and increasing the color scale.\label{fig:sources}}
\label{fig:initial_plot_source_functions}
\end{figure}
\dnew
The overall shape of the three independent source functions $S_{T_0}$, $S_{T_1}$, $S_{T_2}$ is shown in Figure~\ref{fig:sources}. $S_{T_0}$ and $S_{T_2}$ are strongly dominated by terms proportional to $g$ and $g'$, that peak around the time of recombination. Due to the linear color coding of the plots, these dominating terms prevent us from seeing the contribution from reionization, which arises from the second peak of $g$ at late times. This contribution is however important in the final calculation of the CMB spectra especially in the case of $S_{T_2}$, because it determines the polarization spectrum on large angular scales (and in particular the reionization peak). We show it in the lower right panel of Figure~\ref{fig:sources}. Similarly, the term proportional to $e^{-\kappa}$ in $S_{T_0}$\,, which contains most of the early and late Integrated Sachs-Wolfe (ISW) contributions, is not visible in comparison to the dominant terms. The function $S_{T_1}$ only has a term proportional to $e^{-\kappa}$ accounting for the remaining part of the ISW effect, which is clearly visible on the plot for $\tau_\mathrm{rec} \leq \tau \leq  6 \tau_\mathrm{rec} $.

\subsubsection{Where are neural networks most useful?}\label{sub:where}

EBSs perform a sequence of clearly separated tasks or steps in order to compute final observables as a function of input cosmological parameters. As explained in the introduction, we do not want to use NNs to emulate all steps at once, but only one particularly well-suited step, in order to keep the problem simple, the training phase short, and the range of application of each trained version of the NNs as wide as possible.
\dnew
The internal structure of EBSs is well documented in several references, manuals and courses on {\sc cmbfast},  {\sc camb},\footnote{\url{https://lambda.gsfc.nasa.gov/toolbox/tb_cmbfast_ov.cfm} and \url{https://camb.info/}, respectively} and \class{}, and we will not repeat it here. For the purpose of this work, it is sufficient to know that some tasks are faster than others by many orders of magnitude: for instance, the integration of background and thermodynamical quantities, performed at the beginning of EBSs, only requires a few milliseconds.
\dnewnoindent
The two main bottlenecks in EBSs are:
\begin{enumerate}
\item The integration of cosmological perturbations over time that provides all transfer functions (like $F_0(k,\tau)$) and related source functions (like $S_{T_i}(k, \tau)$).
\item The line-of-sight integrals which project the transfer function from Fourier space to harmonic space, like in \fulleqref{eq:angpow delta}.
\end{enumerate}
The slowest between these two steps depends on the context. When cosmologists compute some non-trivial output (e.g.~the $C_\ell$ spectra of cosmic shear or galaxy number count in many redshift bins and with many cross-correlations) for a simple cosmology (e.g.~$\Lambda$CDM), step 2 is slower with current public versions of EBSs. Conversely, when computing simple observables (e.g.~only the CMB spectra and matter power spectrum $P(z,k)$) for a non-trivial cosmological model (e.g.~with massive neutrinos, warm or interacting dark matter, quintessence or modified gravity models with new oscillating degrees of freedom), step 1 is the slowest. In the most frequent situations (e.g.~CMB and $P(k)$ only for $\Lambda$CDM including one massive neutrino), they require a comparable amount of CPU time. Thus, in any case, both steps deserve a deep effort of modernization in order to face the challenges of future survey analyses.
\dnew
Step 1, which we call here the \tquote{perturbation module}, is ideally suited for an NN approach, because it is very difficult to optimize with High Performance Computing (HPC)  techniques, such as massive parallelization or vectorization. In linear cosmological perturbation theory, there is one independent system of ODEs for each wavenumber $k$. Thus the perturbation module features an outer loop over $k$ values, that is easy to parallelize. However, it is difficult to go beyond that, since ODE algorithms are sequential in nature (new time steps depend on previous time steps). Besides, the time needed to integrate the ODEs for each wavenumber increases with $k$. The largest $k$ value in the problem typically requires ${\cal O}(10\%)$ of the total CPU time in the perturbation module. Thus the parallelization of this module cannot scale efficiently above ${\cal O}(10)$ threads, setting an absolute limit on the wall-clock time taken by the EBS, independently of the number of threads. For a simple cosmological model with massive neutrinos, this limit is of the order of one second. We will see that when we emulate the source function calculation with NNs instead of performing the full ODE integration, this times shrinks to a point where step 1 has a negligible contribution to the total EBS execution time (for whatever cosmology and observables).
\dnew
Step 2, which we call here the \tquote{harmonic transfer module}, can instead be tackled with HPC. It consists of a large number of independent integrals, that can even be cut into pieces. Thus there is in principle no limit to the degree of parallelization or vectorization of this module: with some appropriate coding updates, it could always use  the full power of modern CPUs or GPUs. Independently of this brute-force approach, there is also a possibility to reformulate the mathematical transformations employed to calculate the $C_\ell$ spectra from the transfer functions~\cite{Assassi:2017lea,Schoneberg:2018fis}, in order to avoid slowly-converging integrals with quickly oscillating arguments like in \fulleqref{eq:angpow delta}. A third possibility would be to tackle these calculations with a dedicated machine learning technique, distinct from the one discussed in this work due to the different nature of the problem. 
\dnew
In this paper, our purpose is to optimize step 1 (the perturbation module) with an emulator based on NNs, while leaving the optimization of step 2 for future work. This strategy offers considerable advantages:
\begin{itemize}
\item First, the source functions are smoother and closer in shape to analytic approximations than other quantities like the harmonic transfer functions $\Delta_\ell(k)$ or the final spectra $C_\ell$'s. We will show in the next section that this allows to train NNs that reach the required precision level after a surprisingly short training stage. 
\item Second, the source functions do not depend on all of the parameters and aspects of each cosmological model: they are independent of the primordial fluctuation spectrum (and also, when relevant, of the ingredients allowing to estimate non-linear corrections). They simply depend on the parameters of the background cosmology. For instance, in the minimal $\Lambda$CDM model, they only depend on $\{\omega_{\mathrm{b}}, \omega_{\mathrm{cdm}}, H_0, \kappa_{\mathrm{reio}}\}$. This reduced number of parameters speeds up the training and extends the range of validity of the NNs. For instance, there would be no need to retrain the NNs for testing non-minimal inflationary models with a running of the spectral index or with some features in the primordial spectrum.
\item Third, we only emulate a step that takes place before the user needs to specify parameters related to the observables, like window functions describing redshift bins, or parameters modeling some bias or instrumental effects. Thus, a user deciding to switch to a new binning strategy would not need to retrain the NNs.
\item Fourth, this approach is aligned with the needs of the Large Scale Structure community, which tends more and more to use EBSs only for the calculation of transfer and source functions, i.e.~only up to the perturbation module, while using their own codes and approximations (like the Limber approximation~\cite{1953ApJ...117..134L}) 
to compute the final observables. 
\end{itemize}

\subsubsection{Analytical approximations for transfer functions}\label{sub:analapprox}

The CMB source functions depend on two thermodynamical functions of time ($e^{-\kappa}$, $g$), seven transfer functions of time and wavenumber ($F_0$, $F_2$, $G_0$, $G_2$, $\theta_b$, $\phi$, $\psi$) and their derivatives. Here we provide a few analytical approximations to these transfer functions, which can be passed to the NN in order to optimize its efficiency as explained in Section~\ref{sub:network_architecture}.
\dnew
In the baryon-photon tight coupling regime $\kappa' \gg aH$, some of the transfer functions in the previous list remain vanishingly small ($F_2$, $G_0$, $G_2$) or are very smooth ($\phi \simeq \psi$ are constant on super-Hubble scales and smoothly decay on sub-Hubble scales). Thus the functions that are potentially difficult to interpolate or emulate are $F_0$ and $\theta_b$. Due to the tight coupling regime, $\theta_b \simeq \theta_\gamma = 3 F_0'+3 \phi'$. Thus, in order to capture the non-smooth part of the transfer function, we are only interested in analytic approximations for $F_0$, from which all other functions can be calculated. For that we will use the driven oscillator equation that captures the behavior of photon density perturbations deep in the tight coupling limit \cite{Hu:1996mn},
\begin{equation}
	F_0''  + \frac{R'}{1+R} F_0' +k^2 c_s^2 F_0 = -\frac{k^2}{3} \psi + \frac{R'}{1+R} \phi' + \phi''~,
\end{equation}
which involves the baryon-to-photon ratio $R = \frac{3\rho_b}{4\rho_\gamma}$ and the adiabatic photon-baryon sound speed $c_s = (3(1+R))^{-1/2}$. The solution to this equation can be found using the WKB approximation as \cite{Hu:1996mn}
\begin{equation}\label{eq:wkbapprox}
F_0 = -(1+R) \psi + \left[ A \cos\left(k r_s\right) + B \sin\left(k r_s\right)\right]\, \cdot \, \exp\left(-\frac{k^2}{k_D^2}\right)~,
\end{equation}
involving the comoving sound horizon $r_s = \int c_s d\tau$, the coefficients $A$ and $B$ and the diffusion damping wavenumber $k_D$ given in good approximation by 
\begin{equation}
	k_D^{-2} = \frac{1}{6} \int \frac{R^2 + 16/15 (1+R)}{\kappa'(1+R)^2}d\tau~.
\end{equation}
The first term in equation~(2.6) is again a smooth term, while the quickly-varying terms in $F_0$ depend on $\cos(kr_s)$, $\sin(k r_s)$ and $\exp(-k^2/k_D^2)$. Those in the derivatives $F_0'$ and $F_0''$ naturally depend on the very same functions with different and slowly varying coefficients.
\dnew
Well after recombination, the interaction rate $\kappa'$ becomes negligible and photons enter the free-streaming regime. The oscillating component of $F_\ell$ is given by the equation
\begin{equation}\label{eq:dglfreephoton}
	F_\ell' = \frac{k}{2\ell+1}\left[\ell F_{(\ell-1)} - (\ell+1)F_{(\ell+1)}\right]~,
\end{equation}
where we neglected the slowly varying transfer functions ($\theta_b$, $\phi$, $\psi$). Equation \eqref{eq:dglfreephoton} is then solved by
\begin{equation}\label{eq:freeapprox}
	F_\ell(k,\tau) = j_\ell(k\tau)~.
\end{equation}
The very same applies to the polarization perturbations $G_\ell$\,. The oscillatory part of the CMB source functions thus depends on the spherical Bessel functions $j_0(k\tau)$, $j_1(k\tau)$, and $j_2(k\tau)$. Alternatively, we may say that they depend only on $j_1(k\tau)$ and $j_2(k\tau)$, since recurrence relation can give $j_0(k\tau)$ as a function of $j_1(k\tau)$ and $j_2(k\tau)$.

\subsection{Network architecture}
\label{sub:network_architecture}

\subsubsection{Number of networks and output shape}

As described in \fullsecref{sub:where}, we want the network to output the source functions $S_{X}(k,\tau)$ introduced in \fullsecref{sub:equations} for $X \in \{ T_0, T_1, T_2$\}. These functions need to be sampled accurately: we will justify in \fullsecref{sub:Generating_data} that a grid of $\sim 500\times500$ points in $k$ and $\tau$ is necessary to achieve a good precision on the final CMB spectra (this is also the approximate number of values sampled by \class{} when running with default precision). 
\dnew
The first decisions to be made regards the number of networks and the shape of the output. For instance, we could introduce a single network taking $X$ as an input index, or split the problem in three independent networks, one for each different $X$. Given that all three source functions have significantly different shapes in $(k,\tau)$-space (see Figure \ref{fig:sources}), we found that it is more efficient to design a separate network for each source function.
\dnew
We could limit ourselves to these three networks or further divide the problem into independent networks for each $\tau$ and/or each $k$. However, the source functions are continuous in $(k,\tau)$-space, with simple symmetries and regular patterns: thus we expect to maximize overall performance by limiting ourselves to a single network per index $X$. We end up with a total of three independent networks for the calculation of CMB primary anisotropies.
\dnew
There are four distinct ways of predicting the output function of any network: (1) predicting the entire two-dimensional grid $S_{X}(k_i,\tau_j)$ (where $(i,j)$ are the indices of discrete $(k,\tau)$ values), (2) predicting only the one-dimensional grid $S_{X}(k_i,\tau)$ for each $\tau$ passed as an input, (3) predicting one-dimensional grids $S_{X}(k,\tau_j)$ with $k$ passed as in input, or (4) simply predicting single numbers $S_{X}(k,\tau)$ if both $k$ and $\tau$ are passed as an input to the network. 
\dnew
EBSs tend to define an optimal sampling of both $\tau$ and $k$ dynamically, for each cosmological model. However, the possibility to define the sampling dynamically is more important for $\tau$ than for $k$, at least in \class{}. Indeed, the code automatically refines the sampling close to the recombination and reionization times, which are different for each cosmological model. \class{} also defines the $k$-sampling dynamically as a function of the Hubble and sound horizon scales, but dropping this feature is not harmful, because the behavior of the source function in $k$ is more regular and depends less on the thermodynamical input parameters like e.g.~$\kappa_{\mathrm{reio}}$. A fixed $k$ sampling can easily cover all cosmologies.
\dnew
We thus want to avoid setting a fixed grid in $\tau$ space for the network for all cosmologies, since the corresponding sampling around reionization would have to be very fine to cover the narrow peak of reionization for all possible values of $\kappa_\mathrm{reio}$\,. On this basis we choose to exclude options~(1)~and~(3). Option~(4) would require $\mathcal{O}(250 000)$ separate evaluations of the network, and we thus expect a large evaluation overhead for this case. We reserve further exploring ways to circumvent these limitations to future work.
\dnew
Finally, we choose option (3) and thus only target single slices of constant $\tau$ for the output, while supplying the value of $\tau$ at which to evaluate the source function as an input to the network. It means that each of the three networks $\mathrm{NN}_{X}$ with $X \in \{ T_0, T_1, T_2$\} can be seen as a function:
\begin{equation}
\{\tau, \mathrm{cosmological}~\mathrm{input} \} \xmapsto[]{\,\,\,\mathrm{NN}_X\,\,\,} \{ S_X(k_i, \tau) ~\mathrm{for}~\mathrm{each}~ i=1,...,N_k\}~.
\end{equation}
This reduces the size of the output layer $N_k$ to only $\sim 500$ nodes. We let \class{} define the optimal $\tau$ sampling dynamically, not changing anything with respect to the default version of the code. We then iterate over these $\tau$ values to obtain the source function on a predefined grid of $\sim500$ values in $k$ space. In the later steps, \class{} interpolates between these values when computing the CMB spectra. 
\dnew
In the following, we describe how we choose the specific physical inputs and internal architecture of each network.

\subsubsection{Inputs}
\label{subsub:inputs}

It is still a difficult task for each network to predict accurately $\sim$\,500 oscillatory functions of $\tau$ (one for each $k_i$) starting from a set of few cosmological parameters only (namely, the four parameters $\{\omega_{\mathrm{b}}, \omega_{\mathrm{cdm}}, H_0, \kappa_{\mathrm{reio}}\}$ in the $\Lambda$CDM model). We started our investigation with a network input limited to these four parameters and we did not obtain a promising performance: even after several days of training, our networks never reached the required precision.
\dnew
Fortunately, NNs can take redundant input. Since the EBS takes only a few milliseconds to integrate the background and thermodynamical equations before calling the NNs, we have a lot of information to pass to the NNs without any significant computational cost. There is however a trade-off, because passing a larger input vector implies a bigger network size and a potential increase in training and evaluation time. If the information is useful (because it describes some characteristic of the source functions) the accuracy of the NNs increases much faster as a function of the training time, and the tiny increase in evaluation time justifies this efficiency gain. Otherwise, the extra input just makes the NNs heavier without a net efficiency gain.
\dnew
Our investigation of the optimal input consisted of including additional input parameters sequentially, guided by a physical understanding of the source function behavior. Keeping the additions that increased the  efficiency significantly and discarding the others, we converged towards a version of the NNs that reach the desired accuracy after just half a day of training on a single laptop\footnote{To remove any ambiguity, we should stress that the investigations summarized in section \ref{sec:training_strategy} required a lot more computing time, because we actually compared several network architectures in order to select the most efficient one. Once this optimization has been done, training the network in its selected configuration is really as fast as indicated.}, as shown in sections \ref{sec:training_strategy} and \ref{seb:results}.  Here we present only this version.
\dnew
First, let us focus on $S_{T_0}$\,. Looking at \fulleqref{eq:ST0}, one notices that all transfer functions are weighted by the three thermodynamical functions $e^{-\kappa(\tau)}$, $g(\tau)$, and $g'(\tau)$, that define the envelope of $S_{T_0}$ as a function of time. Passing them as input boosts the efficiency of the network. The actual structure, namely the oscillations visible in figure~\ref{fig:initial_plot_source_functions}, arise from the evolution of $F_0(k,\tau)$, $\theta_b(k,\tau)$, and $\theta_b'(k,\tau)$ around decoupling. In section~\ref{sub:analapprox} we explained that this evolution is well approximated (up to slowly-varying coefficients) by the functions $\sin(k r_s(\tau))$ and $\cos(k r_s(\tau))$ multiplied with the envelope $\exp(-k^2/k_D(\tau)^2)$. We can easily evaluate these functions, because the comoving sound horizon $r_s(\tau)$ and damping wavenumber $k_D(\tau)$ are computed within the thermodynamical module of \class{}. These functions give the network a basis to adjust and modify in order to match the true source functions; thus they are extremely beneficial to the performance of the network. Finally, we found that the training efficiency also benefits from explicitly passing some characteristic times and scales that are directly related to features in the source function, and come with basically no additional computational cost and only a small amount of extra weights. These are the conformal time at recombination and reionization, $\tau_{\mathrm{rec}}$ and $\tau_{\mathrm{reio}}$\,, the comoving and physical sound horizon at recombination, $r_s^\mathrm{rec}$ and $d_s^\mathrm{rec}$, the comoving and physical angular diameter distance to recombination,  $r_a^\mathrm{rec}$ and $d_a^\mathrm{rec}$, and the comoving damping scale at recombination, $r_D^\mathrm{rec}$. 
\dnew
For the network predicting $S_{T_1}$\,, the inputs are largely the same, except that the visibility function $g(\tau)$ and its derivative are omitted as they are not present in \fulleqref{eq:ST1}. Note that the difference $\psi-\phi$ is primarily sourced through the Einstein equation involving the total shear, which is dominated by the photon shear. This shear is directly related to $F_0$ and its time derivatives -- c.f. equation (\ref{eq:photonboltzmann}) --, and thus mainly given by a superposition of $\sin(k r_s(\tau))$ and $\cos(k r_s(\tau))$ multiplied by the damping envelope $\exp(-k^2/k_D(\tau)^2)$. Hence the same $k$-dependent input can be passed to the $S_{T_0}$ and $S_{T_1}$ networks.
\dnew
For the $S_{T_2}$ network, \fulleqref{eq:ST2} shows that we should keep the visibility function $g(\tau)$ and omit $g'(\tau)$ and $e^{-\kappa(\tau)}$. Physically, this source function accounts for the generation of polarization when photons scatter off electrons that are surrounded by a quadrupole temperature anisotropy, around the time of both recombination and reionization. Rescattering at reionization is responsible for the \tquote{reionization bump} in the $C_\ell^{EE}$ polarization spectrum. Thus, the network should be able to accurately predict the late-time structure in $S_{T_2}$\,, which is clearly visible in the lower right panel of Figure~\ref{fig:sources}. When we provide analytical approximations to the network, we should in  principle pass the tight-coupling solution to account for structures around recombination and the free-streaming solution to account for structures around reionization. These two sets of approximations are given by oscillatory functions that have different frequencies and phases. We handle this by passing a combined function that switches from the  cosine/sine solutions to the spherical Bessel functions at some intermediate time, using a smooth step function:
\begin{subequations}
	\begin{align}
		&\cos_\mathrm{in} = \mathrm{step}(\tau) \cdot j_1(k \cdot [\tau - \tau_{\mathrm{rec}}]) + [1-\mathrm{step}(\tau)] \cdot \cos(k r_s(\tau))\\
		&\sin_\mathrm{in} = \mathrm{step}(\tau) \cdot j_2(k \cdot [\tau - \tau_{\mathrm{rec}}]) + [1-\mathrm{step}(\tau)] \cdot \sin(k r_s(\tau))
	\end{align}
	with the step function centered at $4 \tau_{\mathrm{rec}}$ defined as 
	\begin{align}
		\mathrm{step(\tau)} = \frac{1}{2}\left[\tanh\left(\frac{\tau - 4 \tau_{\mathrm{rec}}}{100}\right)+1\right]~.
		\label{eq:step}
	\end{align}
\end{subequations}
Note that the spherical Bessel functions $j_1(x)$ and $j_2(x)$ are rescaled as described in the next paragraph.
Passing these combined functions gives much better performances than passing separately the cosine/sine and Bessel functions.	
\dnew
Not only the functional form, but also the overall amplitude of the input is important for the network. Indeed, it is customary to rescale the input and output vectors of an NN before training it, to ensure that all  terms in these vectors share the same typical order of magnitude -- for instance, order one. This is already the case for our cosine and sine functions, as well as $\exp(-k^2/k_D^2)$ and $\exp(-\kappa)$. We rescale the Bessel functions to a peak value of one before combining them with the cosine and sine.
\dnew
Additionally, we rescale the conformal times $\tau$, $\tau_{\mathrm{rec}}$ and $\tau_{\mathrm{reio}}$. To keep their relative meaning to each other intact, we chose to apply a logarithm to these quantities. To rescale $g(\tau)$ and $g'(\tau)$, we divide them by the mean of their maximum value, calculated across the whole training set, such that all the rescaled functions peak at a value of order one. For other individual quantities, like the cosmological parameters and the characteristic scales, we employ the method described in the following. 
\dnew
For each individual parameter, we can compute a mean and a standard deviation over the training set:
\begin{equation}
\bar{x} = \frac{1}{N}\sum_{\mathrm{set}}x_i~, 
\quad
\sigma(x) = \left( \frac{1}{N}\sum_{\mathrm{set}}(x_i-\bar{x})^2 \right)^{1/2}~, \:\:\: x_i \in \mathrm{set}~.
\end{equation}
Then we define the following affine rescaling of $x$:
\begin{equation}
x \longmapsto  \tilde{x} = \frac{x - \bar{x}}{\sigma(x)}~,
\end{equation}
ensuring that most $|\tilde{x}|$ values are of order one.
\dnew
The same procedure is carried out for the source functions: for each $X$, we compute the average $S_X(k,\tau)$ across a grid in $(k,\tau)$ space and across the training set. We then compute the standard deviation for this average and rescale  $S_X(k,\tau)$  by the same affine function as for individual parameters. Thus the networks is actually trained to predict the rescaled functions $\tilde{S}_X(k, \tau)$, that can be readily transformed back to the physical $S_X(k,\tau)$ with the inverse affine function. All the coefficients of the affine functions used to rescale input and output quantities are of course stored together with the NNs.
\dnew
We have now justified all the characteristics of our NN input quantities, which are summarized in Table \ref{tab:inputs}. Additionally, they are shown within the full architecture of the networks in figure \ref{fig:network_architecture}.

\renewcommand{\arraystretch}{1.2}
\setlength{\tabcolsep}{10pt}
\begin{table}[ht]
	\centering
	\begin{tabular}{cc|cc|cc|cc}
		\toprule
		 \multicolumn{2}{c|}{\thead{Times and\\thermal evolution}} & \multicolumn{2}{c|}{\thead{Cosmological\\basis}} & \multicolumn{2}{c|}{\thead{Characteristic\\scales}} & \multicolumn{2}{c}{\thead{$k$-dependent\\approximations}} \\ \midrule \rule{0pt}{16pt}
		
		$g(\tau)/\overline{g_\mathrm{max}}$ & 1 & $\tilde{\omega}_{\mathrm{b}}$ & 1 & $\qquad \tilde{r}_s^{\mathrm{rec}}$ & 1 & $\cos_\mathrm{in}(k_i r_s(\tau))$ & 568 \\
		$g'(\tau)/\overline{g'_\mathrm{max}}$ & 1 & $\tilde{\omega}_{\mathrm{cdm}}$ & 1 & $\qquad \tilde{d}_s^{\mathrm{rec}}$ & 1 & $\sin_\mathrm{in}(k_i r_s(\tau))$ & 568 \\
		$e^{-\kappa(\tau)}$ & 1 & $\tilde{H}_0$ & 1 & $\qquad \tilde{r}_a^{\mathrm{rec}}$ & 1 &  $e^{-(k_i/k_D(\tau))^2}$ & 568 \\
		$\log_{10}\tau_{\mathrm{rec}}$ & 1 & $\tilde{\kappa}_{\mathrm{reio}}$ & 1 & $\qquad \tilde{d}_a^{\mathrm{rec}}$ & 1 &  & \\
		$\log_{10}\tau_{\mathrm{reio}}$ & 1 &  &  & $\qquad \tilde{r}_d^{\mathrm{rec}}$ & 1 &  & \\
		$\log_{10} \tau$ & 1 &&&&&&\\
		\bottomrule
	\end{tabular}
	\caption{Inputs of the network and shape (dimension) of each input quantity. The precise size of the $k$-dependent input is governed by the $k$-sampling used throughout our work, which is detailed in \fullsecref{sub:Generating_data}.\label{tab:inputs}}
\end{table}
\setlength{\tabcolsep}{6pt}
\renewcommand{\arraystretch}{1}

\subsubsection{Layers and connections} 
\label{subsub:layers_and_connections}

\begin{figure}[t!]
	\centering
	\includegraphics[width=0.58\linewidth]{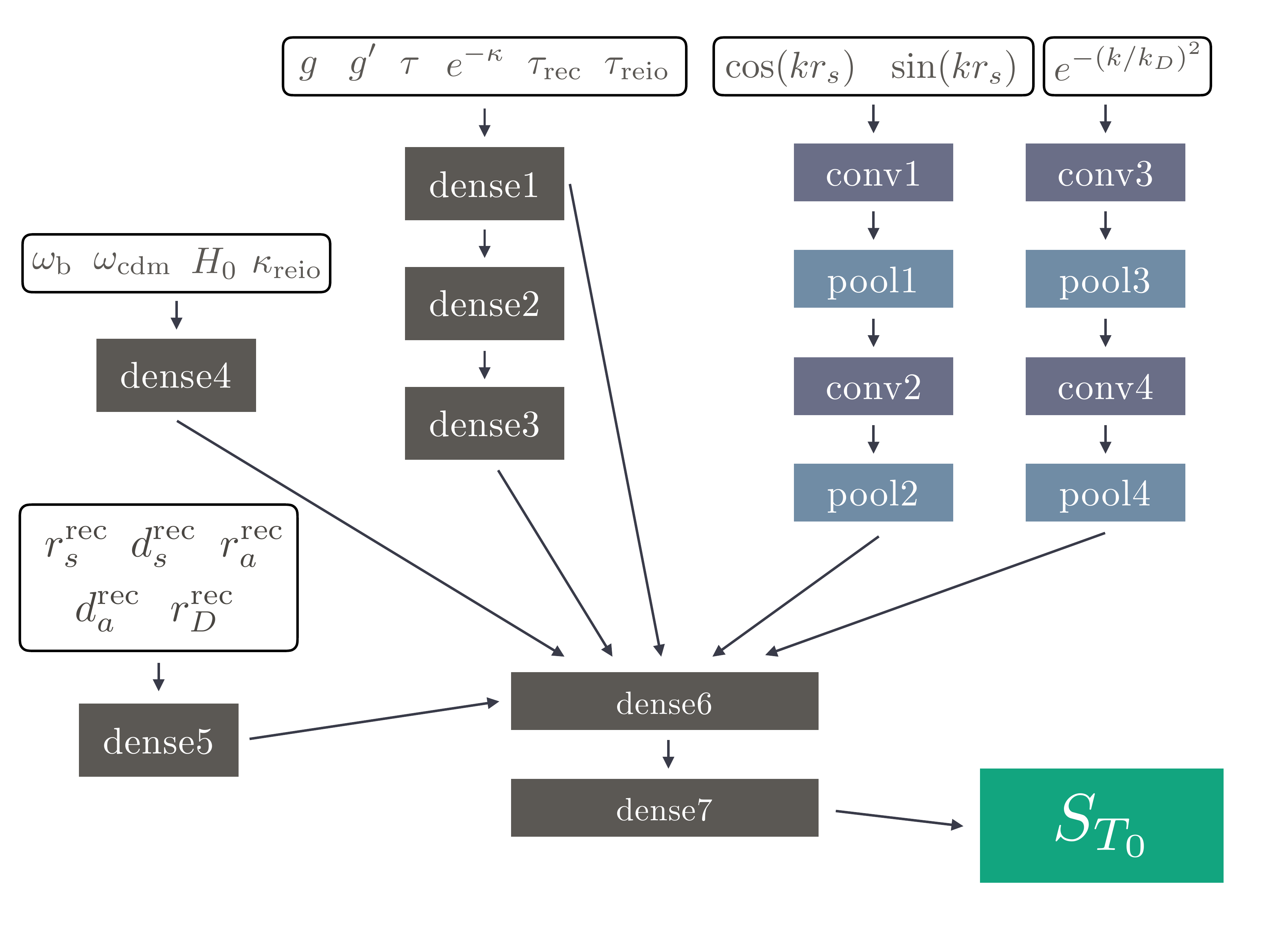}\\
	\includegraphics[width=0.58\linewidth]{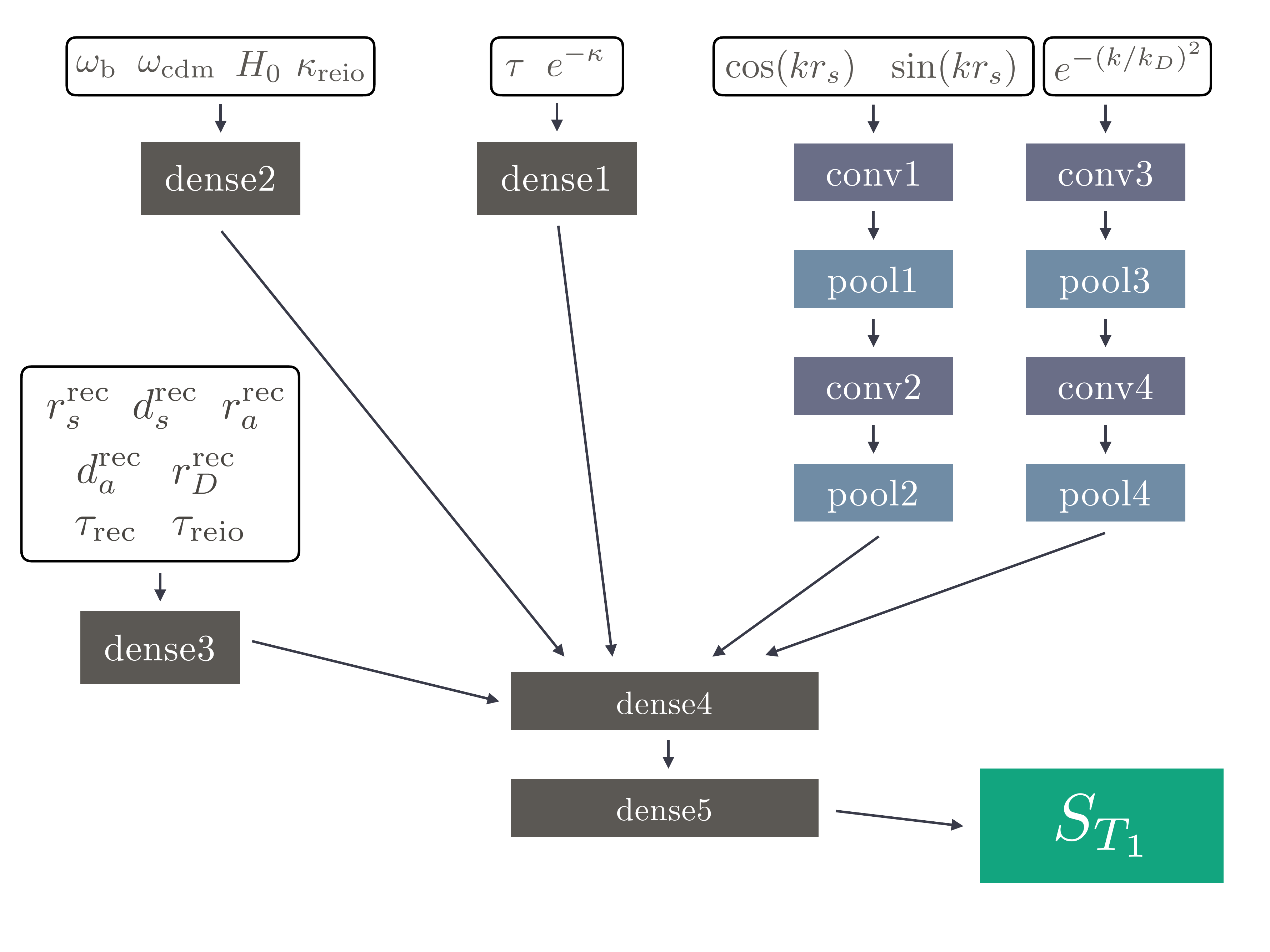}\\
        \includegraphics[width=0.58\linewidth]{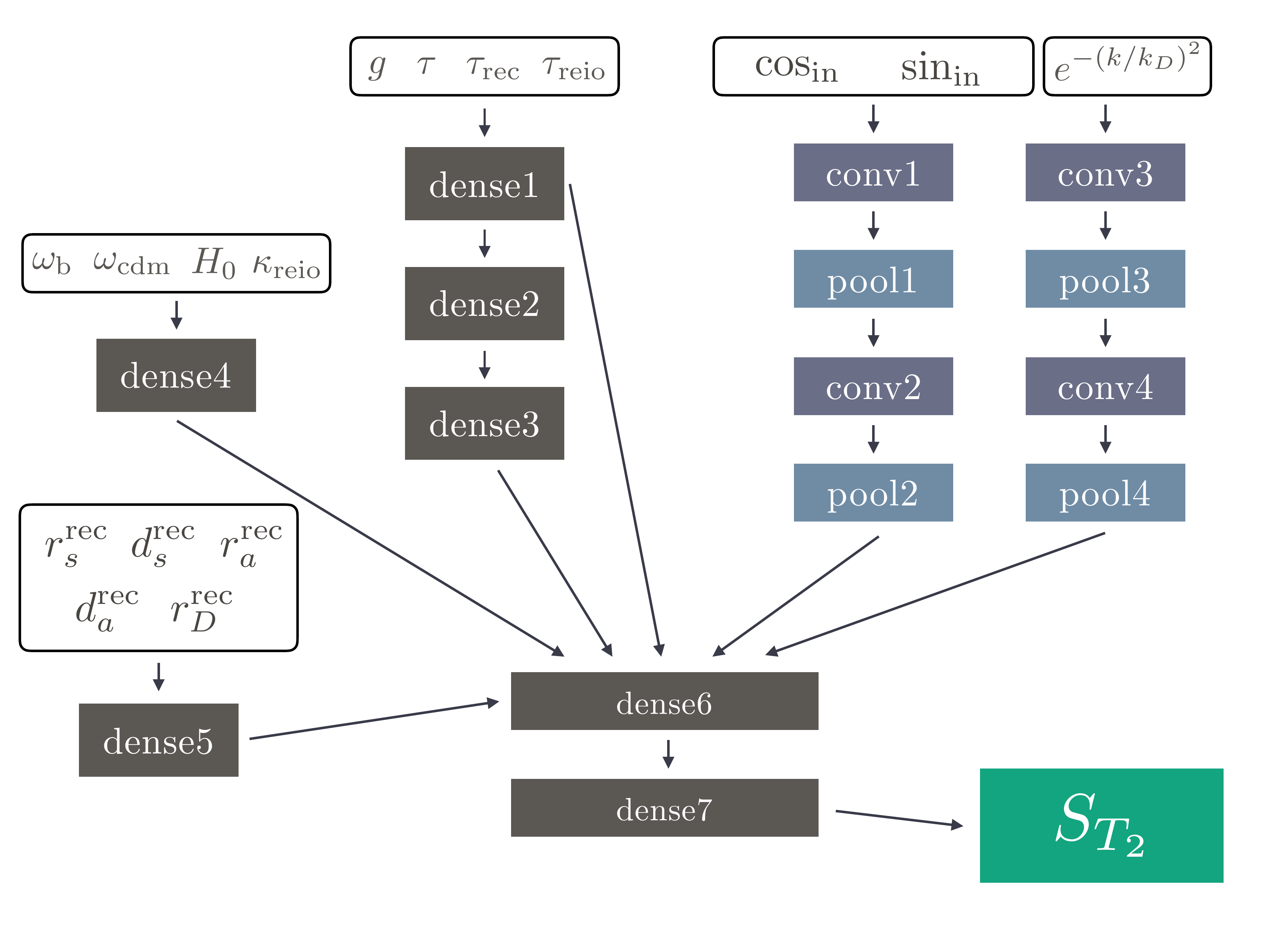}
	\caption{Network architecture of the three networks. White boxes show input quantities, dark boxes stand for dense (fully connected) layers, and blue boxes for convolutional layers.}
	\label{fig:network_architecture}
\end{figure}

\begin{table}[t]
	\begin{minipage}[b]{.5\linewidth}
		\centering
		\begin{tabular}{lccc}
		\toprule
		\textbf{Layer} & \textbf{Activ.} & \textbf{Shape} & \textbf{Params.} \\ 
		\midrule
		dense1 & LReLU & 192 & 1K \\
		dense2 & LReLU & 56 & 10K \\
		dense3 & LReLU & 138 & 8K \\
		\midrule
		dense4 & LReLU & 196 & 1K \\
		\midrule
		dense5 & LReLU & 24 & 0.1K \\
		\midrule
		dense6 & LReLU & 289 & 143K \\
		dense7 & LReLU & 154 & 44K \\
		\midrule
		conv1 & ReLU & 95 x 15 & 0.2K \\
		pool1 & - & 47 x 15 & - \\
		\midrule
		conv2 & ReLU & 16 x 9 & 0.5K \\
		pool2 & - & 8 x 9 & - \\
		\midrule
		conv3 & ReLU & 114 x 7 & 0.1K \\
		pool3 & - & 57 x 7 & - \\
		\midrule
		conv4 & ReLU & 19 x 7 & 0.3K \\
		pool4 & - & 9 x 7 & - \\
		\midrule
		$S_{T_0}$ & linear & 568 & 197K \\ 
		\midrule
		\multicolumn{3}{l}{Total trainable parameters} & \textbf{407K}\\
		\bottomrule
		\end{tabular}
	\caption{$S_{T_0}$ network architecture.}
	\label{tab:ST0_network}
	\end{minipage}%
	\begin{minipage}[b]{.5\linewidth}
		\centering
		\begin{tabular}{lccc}
		\toprule
		\textbf{Layer} & \textbf{Activ.} & \textbf{Shape} & \textbf{Params.} \\ 
		\midrule
		dense1 & LReLU & 73 & 0.4K \\
		dense2 & LReLU & 154 & 11K \\
		dense3 & LReLU & 76 & 12K \\
		\midrule
		dense4 & LReLU & 39 & 0.2K \\
		\midrule
		dense5 & LReLU & 14 & 0.1K \\
		\midrule
		dense6 & LReLU & 230 & 66K \\
		dense7 & LReLU & 245 & 57K \\
		\midrule
		conv1 & ReLU & 142 x 6 & 0.1K \\
		pool1 & - & 71 x 6 & - \\
		\midrule
		conv2 & ReLU & 36 x 6 & 0.1K \\
		pool2 & - & 18 x 6 & - \\
		\midrule
		conv3 & ReLU & 95 x 8 & 0.1K \\
		pool3 & - & 47 x 8 & - \\
		\midrule
		conv4 & ReLU & 16 x 6 & 0.3K \\
		pool4 & - & 8 x 6 & - \\
		\midrule
		$S_{T_1}$ & linear & 568 & 181K \\ 
		\midrule
		\multicolumn{3}{l}{Total trainable parameters} & \textbf{328K}\\
		\bottomrule
		\end{tabular}
	\caption{$S_{T_2}$ network architecture.}
	\label{tab:ST2_network}
	\end{minipage}
\end{table}

\begin{table}
	\centering
	\begin{tabular}{lccc}
	\toprule
	\textbf{Layer} & \textbf{Activ.} & \textbf{Shape} & \textbf{Params.} \\ 
	\midrule
	dense1 & LReLU & 98 & 0.3K \\
	\midrule
	dense2 & LReLU & 17 & 0.1K \\
	\midrule
	dense3 & LReLU & 29 & 0.2K \\
	\midrule
	dense4 & LReLU & 50 & 9K \\
	dense5 & LReLU & 44 & 2K \\
	\midrule
	conv1 & ReLU & 41 x 6 & 0.2K \\
	pool1 & - & 20 x 6 & - \\
	\midrule
	conv2 & ReLU & 4 x 9 & 0.4K \\
	pool2 & - & 2 x 9 & - \\
	\midrule
	conv3 & ReLU & 41 x 17 & 0.3K \\
	pool3 & - & 20 x 17 & - \\
	\midrule
	conv4 & ReLU & 4 x 5 & 0.5K \\
	pool4 & - & 2 x 5 & - \\
	\midrule
	$S_{T_1}$ & linear & 568 & 26K \\ 
	\midrule
	\multicolumn{3}{l}{Total trainable parameters} & \textbf{38K}\\
	\bottomrule
	\end{tabular}
\caption{$S_{T_1}$ network architecture.}	
\label{tab:ST1_network}
\end{table}
After several optimization tests, we converged to a network architecture that can be seen in figure \ref{fig:network_architecture} and tables \ref{tab:ST0_network} to \ref{tab:ST1_network}. Here we justify the overall structure of our NNs. The methodology employed to optimize the hyperparameters of the NNs (e.g.~the size of each layer or the learning rate) will be detailed in \fullsecref{sec:training_strategy}. 
\dnew
The central piece in the network architecture is comprised of two final dense (i.e.~fully connected) layers, called {\tt dense6}, {\tt dense7} for $S_{T_0}$ and $S_{T_2}$\,, and {\tt dense4}, {\tt dense5} for $S_{T_1}$\,. These are eventually connected to the output of shape $568$. The layers are fed by separate channels that pre-process the information coming from the different kinds of input.
\dnew
The cosmological parameters and the characteristic scales are already condensed information, and thus require only a single shallow layer before connecting to the final dense part of the network.
\dnew
Since the network processes slices of constant $\tau$ rather than the full grid, all thermodynamical functions of conformal time are just single numbers to it. Together with the input value of $\tau$, these numbers are very important for determining the source function's envelope, and thus have a deep structure of hidden dense layers connecting them to the central dense layer. Additionally, we employ skip-connections from the first dense layer to the central dense layer, as displayed in figure \ref{fig:network_architecture}. This allows the network to keep some direct information about the shape of the visibility function and other time dependent inputs, while also allowing it to construct the envelope more abstractly using the added depth in dense layers.
\dnew
According to our sampling strategy which is detailed in section \ref{sub:Generating_data}, the $k$-dependent approximations are sampled over 568 values of $k$ for each time step. To efficiently couple these inputs to our network, we employ convolutional layers in conjunction with pooling layers. This approach is well established for convolutional neural networks (CNNs) used e.g.~in image recognition tasks. This allows us to drastically decrease the amount of required neurons, as opposed to fully connecting the input using dense layers. Finally, these deep convolutional networks are appended to the input of the central layer.
\dnew
Note that the $S_{T_1}$ network was chosen to have an order of magnitude fewer weights than the other two. During evaluation it became apparent that $S_{T_1}$ has an almost negligible impact on the final temperature anisotropy spectrum of the CMB, and thus some accuracy can be sacrificed in the interest of training and evaluation speed.
\dnew
Concerning the activation functions in each layer, 
we found that the rectified linear unit (ReLU) and the leaky ReLU (LReLU), defined as
\begin{equation}
	\text{ReLU}(x) = \max(0,x) \qquad  
	\text{and} \qquad  
	\text{LReLU}(x) = 
		\begin{cases}
		x &\quad \text{if } x \geq 0\\
		\beta x &\quad \text{if } x < 0\\
		\end{cases}
		, \:\:\:\: 0 < \beta < 1
		\label{eq:ReLU}
\end{equation}
give the best performances using $\beta=0.3$.
\dnew
Both regularization and dropout layers actually proved to worsen the network accuracy, and therefore are not employed in the final version. We believe that our networks do not suffer from the consequences of overfitting or error-prone inputs and thus do not improve using these two strategies.

\section{Training strategy}
\label{sec:training_strategy}

The goal of the work reported in this section is to find the network architecture and the hyperparameter values that lead to the most efficient training stage in terms of speed and precision. This is a computationally expensive investigation, since it involves parallel training of many network instances. We should stress that this heavy computational effort is meant to be done only once, or at least, relatively rarely. Indeed, once the optimal network settings have been found, retraining the network for the purpose of including new cosmological parameters or an extended range in parameter space can be done with the same hyperparameters. Thus it can be achieved rapidly (we will see that each training requires typically half a day on four cores).
\dnew
To create and manipulate NNs, we use the open source platform TensorFlow \cite{tensorflow2015-whitepaper} through the Keras API \cite{chollet2015keras}. The multicore nature of the compute cluster allows to train several hundred realizations of a network simultaneously.

\subsection{Generating data for training and validation} 
\label{sub:Generating_data}

The network's performance will be optimal within the boundaries of our training data set. Thus, before training the network, we have to set these boundaries to encompass all cosmologies of interest. As this is a proof-of-concept paper, we start by focusing on $\mathrm{\Lambda CDM}$ models including a massive neutrino species with a mass of $0.06\,\mathrm{eV}$: this is the baseline model of the Planck CMB survey \cite{Ade:2013zuv}. As stated previously, and exploited by the design of our network, only four of the six free parameters of that model have an effect on the shape of the source functions: $\{\omega_{\mathrm{b}}, \omega_{\mathrm{cdm}}, H_0, \kappa_{\mathrm{reio}}\}$. We train on a hypercube of side $\mu \pm 5\sigma$ for each parameter, where $(\mu,\sigma)$ are the mean and standard deviation of the posterior of Planck 2018 CMB TT,TE,EE+lowE+lensing+BAO data \cite{Aghanim:2018eyx}. In this four-dimensional hypercube we use Latin hypercube sampling (LHS) \cite{10.2307/1268522,10.2307/2291282} to generate 10~000 models to actively train on, plus 2~000 models for network performance validation during training, 100 models for hyperparameter optimization, and 100 models for the final performance test after training. Note that all four sets are completely independent from each other. 
\dnew
We call \class for each of the above models, to compute and store the inputs (thermodynamic and background quantities) and targeted outputs (source functions) of the NN. This step is only required once before any training to initialize the respective sets.
\dnew
For the $k$ grid on which we store the source functions and train the networks, we select the one produced by \class for the Planck 2018 best-fit cosmology. \class chooses a $k$ sampling that is optimized to track features of the source functions well and has increased density around $k$ values particularly important for the CMB anisotropies. The best-fit cosmology sampling also decides the final output size of our network (568).
\dnew
Even if our final network will not involve any $\tau$-grid, for the purpose of training, we need to store $\tau$-dependent functions at discrete values of conformal time. We use a $\tau$-grid very similar to the one defined by \class, but slightly modified to suit our needs. Most importantly, we reduce the sampling density in regions where no structure is present in the source functions, and increase it where accuracy is desirable, especially around $\tau_{\mathrm{rec}}$ and $\tau_{\mathrm{reio}}$. Note that the latter will usually vary more than the former, and we thus choose to cover the full range ($\mu\pm5\sigma$) of possible $\tau_{\mathrm{reio}}$ values with our increased sampling. As shown in section \ref{sub:results_source_functions} and visible in the lower right panel of figure \ref{fig:initial_plot_source_functions}, $S_{T_2}$ receives an important contribution from photon rescattering around the time of reionization. 

\subsection{Training procedure} 
\label{sub:training_procedure}

We have fixed a few quantities according to the results of some preliminary tests. We found that the Adam optimizer \cite{2014arXiv1412.6980K}, which is based on gradient descent, with its default learning rate of $l = 10^{-3}$ works best for our purposes. For the activation function, our tests showed that the ReLU and LReLU described in \fulleqref{eq:ReLU} are optimal. As a loss function we select a mean squared error given by
\begin{equation}
	L(\tau)~=~\sum_{i=1}^{N_k}~(S_{\mathrm{true}}(k_i,\tau)~-~S_{\mathrm{pred}}(k_i,\tau))^2~.
\end{equation} 
Next, we must adjust two parameters governing the duration and the organization of the NN training sequence: the batch size and the amount of epochs. 
\dnew
A basic network training pipeline starts by passing input to the network and letting it make a prediction with which the loss can be computed. Without batching, this is done for each input vector sequentially. In our case, this would mean that the network adjusts its weights after evaluating the source function over a single slice with fixed $\tau$. With batching, the network receives multiple input data, makes predictions for all of them, and calculates a loss for all predictions combined. Only after this step it adjusts its weights accordingly. This is standard practice in training NNs due to performance benefits. In our case, we choose the batch size to be equal to the length of our $\tau$ grid, such that each batch is comprised of all the input and output values corresponding to one cosmology.
\dnew
The amount of epochs defines how often the network trains on the same data. More specifically, choosing $n$ epochs means that the network gets a batch of the specified batch size, calculates the loss and adjusts its weights, feeds in the \emph{same} batch of inputs, calculates newly updated weights, and so on for $n$ times. An increase in the number of epochs directly causes an increase in training time. It is important to note that $S_{T_0}$ has the biggest influence on the temperature anisotropy spectrum, while $S_{T_1}$ and $S_{T_2}$ only play subordinate roles. However, the polarization spectrum depends exclusively on $S_{T_2}$ since $S_P = \sqrt{6} S_{T_2}$. For this reason, we shorten the training time for $S_{T_1}$ as compared to $S_{T_0}$. Conversely, because of the complexity of the shape of $S_{T_2}$, we increase the training time to obtain more precision. Thus, we chose to train each $S_{T_0}$ network for 15 epochs, each $S_{T_1}$ network for 3 epochs and each $S_{T_2}$ network for 45 epochs.
\dnew
Even after fixing all these quantities, the network architecture requires the determination of about 20 hyperparameters which are \emph{a priori} unknown. These include the sizes of the different dense layers and three parameters specific to each convolutional layer; the kernel size specifying the size of the window each node of the layer is connected to, the stride representing the distance the window jumps when scanning the input, and the amount of filters representing the number of different features each layer is able to learn. Additionally, one has to set the window size of the max pooling layers. 
\dnew
For all these hyperparameters one can make some educated guesses about the orders of magnitude in which their optimum lies. However, finding the optimal combination requires drawing each hyperparameter from these ranges multiple times, which we do according to a logarithmic prior. After drawing $\mathcal{O}(500)$ random hyperparameter samples, each network is trained and the history of its validation loss is recorded. Once training has finished, the final validation loss represents a criterion to measure the network performance.
\dnew
While the network only outputs the source functions, the final quantities of interest are the CMB temperature or polarization spectra $C_\ell^{XY}$. A small loss on the source functions means that they are overall very accurate in $(k,\tau)$-space, but not necessarily that we have trained the network optimally in order to achieve a given precision on the $C_\ell^{XY}$'s.
\dnew
Thus one could try to design a strategy such that the network would get information about the accuracy of the CMB spectra during training. Naively one can think to design a custom loss function that calculates the error directly on the $C_\ell^{XY}$'s, therefore steering the training in a direction where this error is minimized. Unfortunately, this is fundamentally impossible. Indeed, the training is based on a gradient descent of the loss function: at each training step, a gradient is calculated and a step is taken in the steepest descent direction. Thus the neural net needs to be able to calculate this gradient. Using an external function or program like \class to transform the output of the network in some non-trivial way unaccessible to the network (namely, into $C_\ell^{XY}$'s) would invalidate the gradient descent approach. We tried to circumvent this by introducing coefficients in the loss function such that regions having more influence on the CMB spectra have more weight, but this approach did not lead to any improvement. We thus devised a different strategy: we stick to a source-function-based loss during each network training, but we select the best network instance on the basis of the CMB spectra accuracy.
\dnew
Our selection of the best network out of our $\mathcal{O}(500)$ instances proceeds along two steps. First, we rank the network by their final validation loss, and keep only the 30 top ones. Second, for these top networks we calculate the CMB spectra inferred from the network output source functions for each of the 100 cosmologies of our hyperparameter optimization set. Finally we evaluate the relative error on the final anisotropies according to
\begin{equation}
	\text{error} = \sum_{\ell} \left( \frac{C_{\ell}^{\text{\class-net}}-C_{\ell}^{\text{\class-full}}}{C_{\ell}^{\text{\class-full}}} \right)^2~.
\end{equation}
and pick the network instance giving the smallest relative error.
\dnew
With final hyperparameter settings and on four CPU cores on the RWTH Compute Cluster (Intel Xeon Platinum 8160 processors), the network training (wall-clock) time is respectively 3h, 0.6h and 10h for $S_{T_0}$, $S_{T_1}$  and $S_{T_2}$\,. Compared to the $S_{T_0}$ network, the $S_{T_2}$ network takes more time to train due to our choice of increasing the number of epochs, while the $S_{T_1}$ network takes less time to train since it has the smallest number of epochs and the smallest size. 
\section{Results}
\label{seb:results}

We implemented the three neural networks that we built up in the previous section inside \class, in such way that the user can choose to compute the CMB spectra in two ways: in \tquote{\class-full} mode (getting the source function for the integration of perturbation equations) or in \tquote{\class-net} mode (getting the source function from the neural network). Since our neural networks are developed in python, the python wrapper {\tt classy.pyx} was the most natural place to implement the network evaluation.
\newpage
\subsection{Accuracy of the source functions}
\label{sub:results_source_functions}

\begin{figure}[t]
	\centering
	\begin{subfigure}[t]{.62\textwidth}
		\centering
		\includegraphics[width=\linewidth]{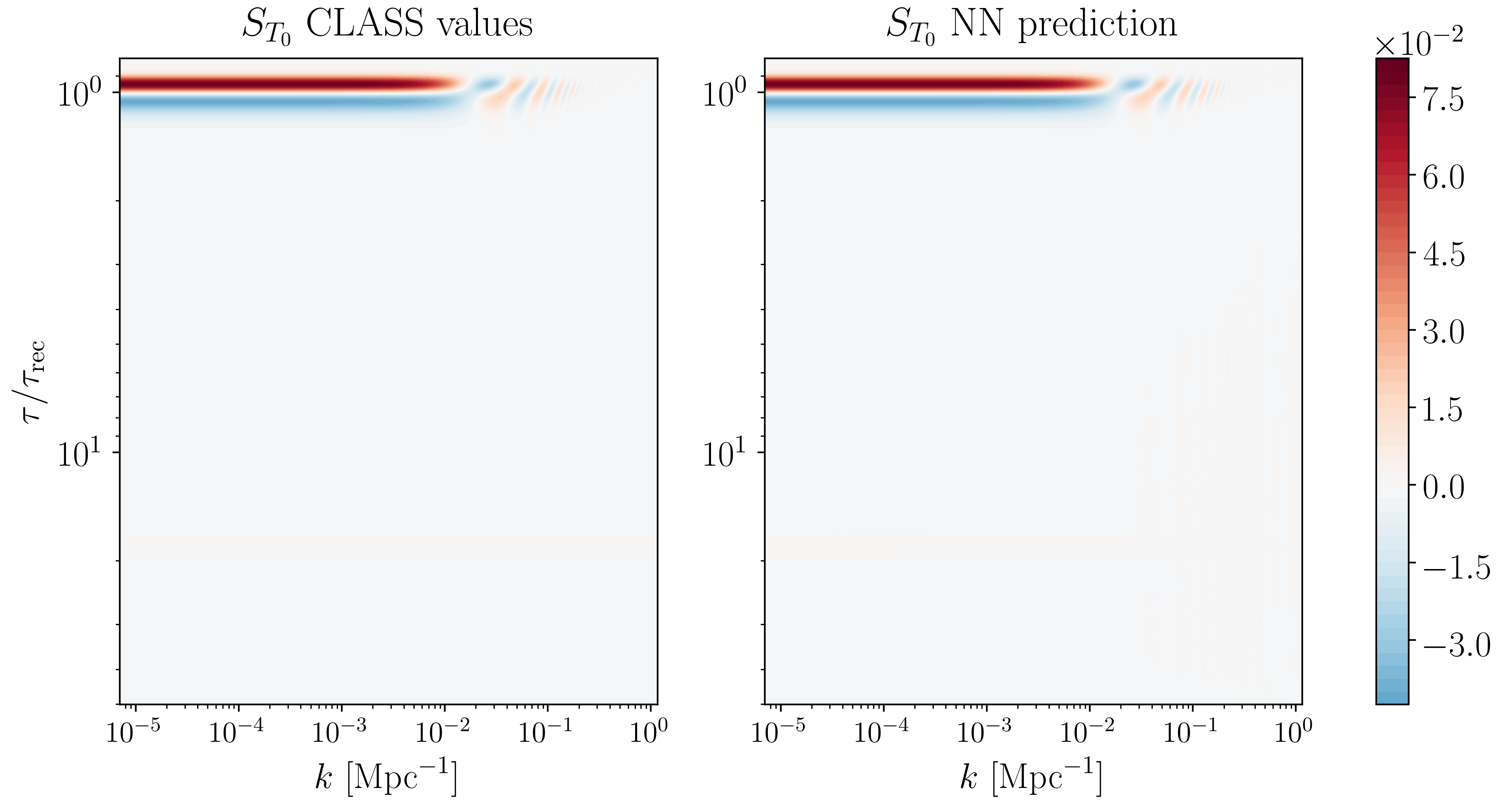}
	\end{subfigure}%
	\begin{subfigure}[t]{.38\textwidth}
		\centering
		\includegraphics[width=\linewidth]{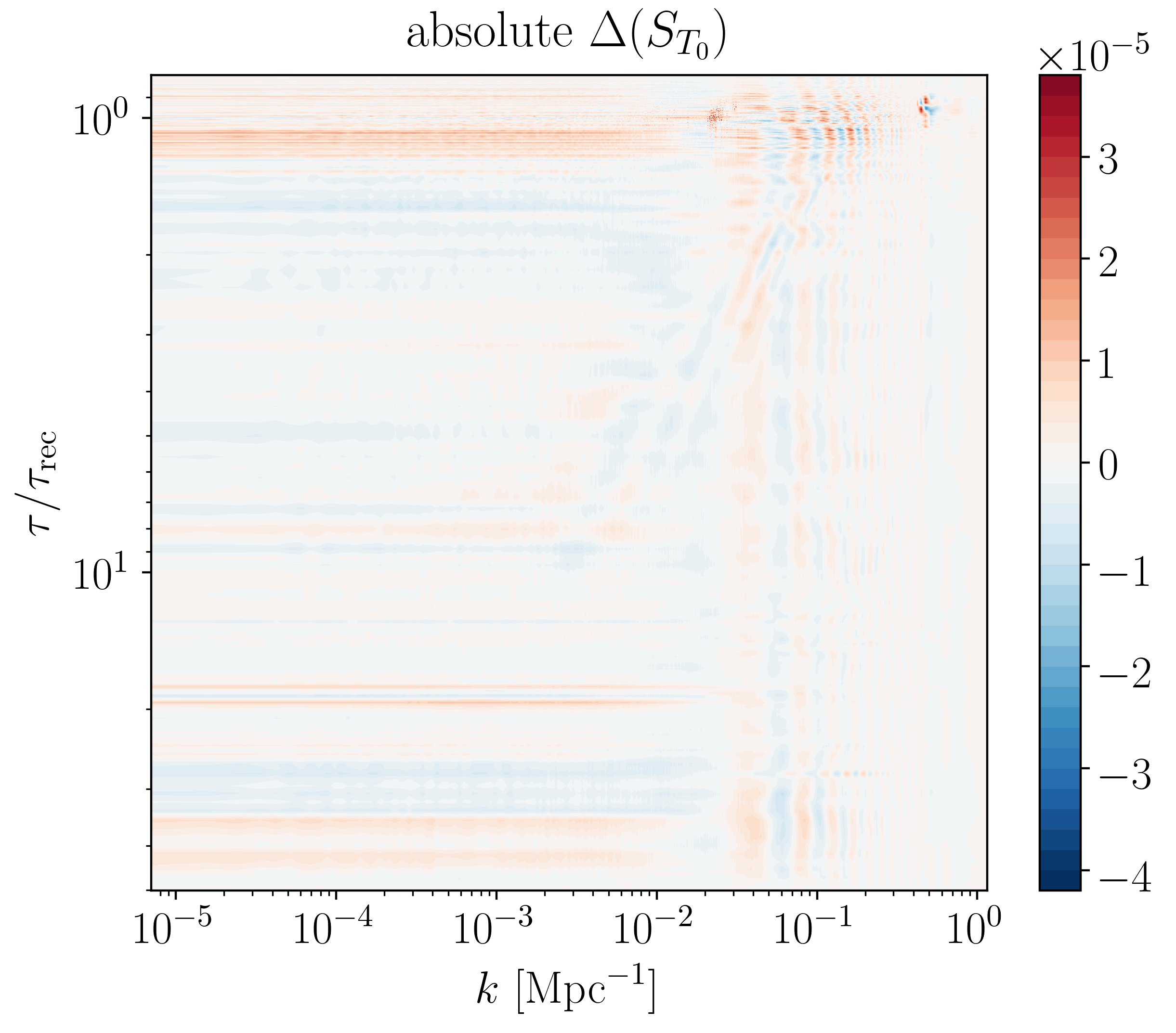}
	\end{subfigure}
	\centering
	\begin{subfigure}[t]{.61\textwidth}
		\centering
		\includegraphics[width=\linewidth]{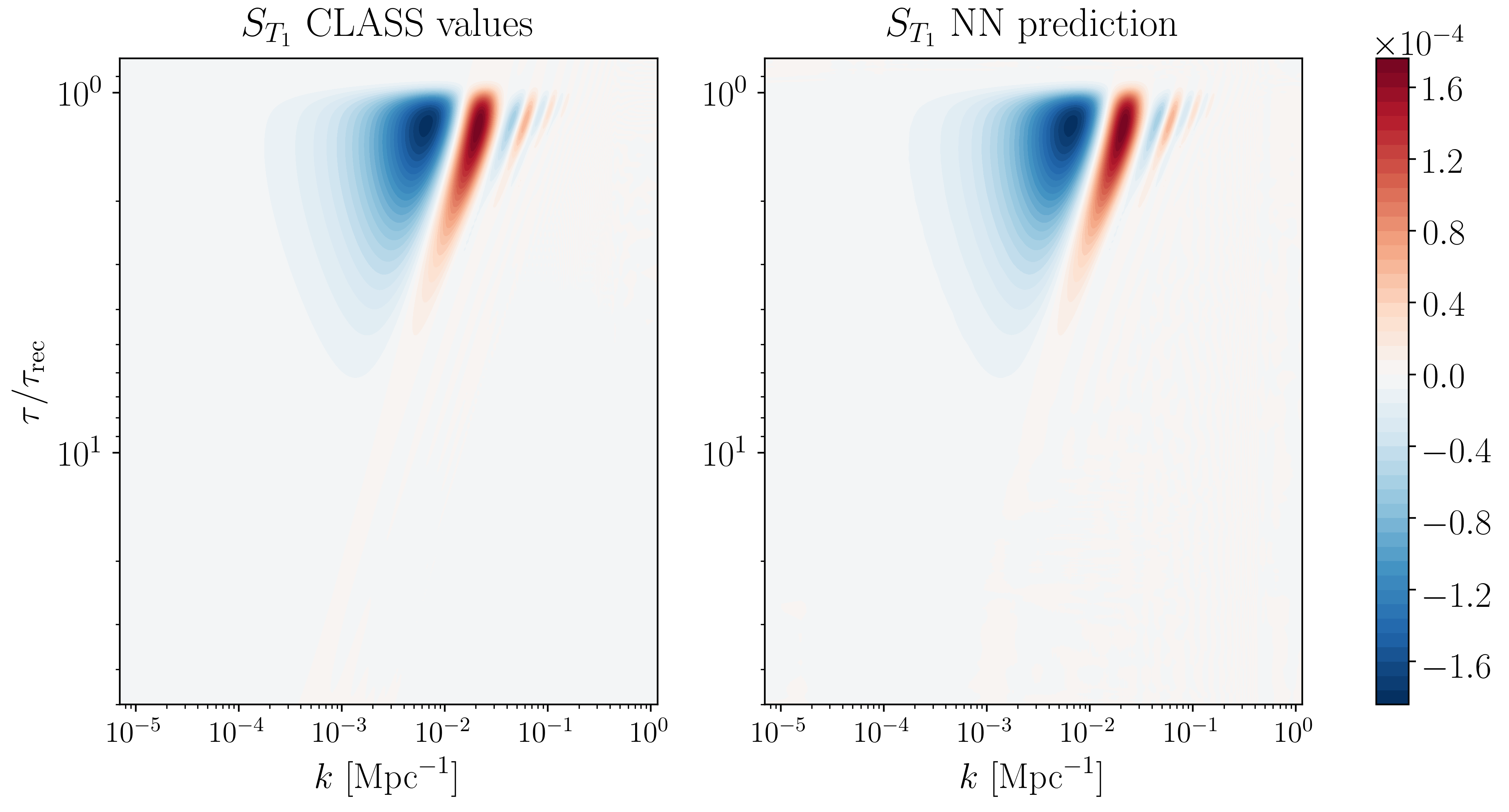}
	\end{subfigure}%
	\begin{subfigure}[t]{.39\textwidth}
		\centering
		\includegraphics[width=\linewidth]{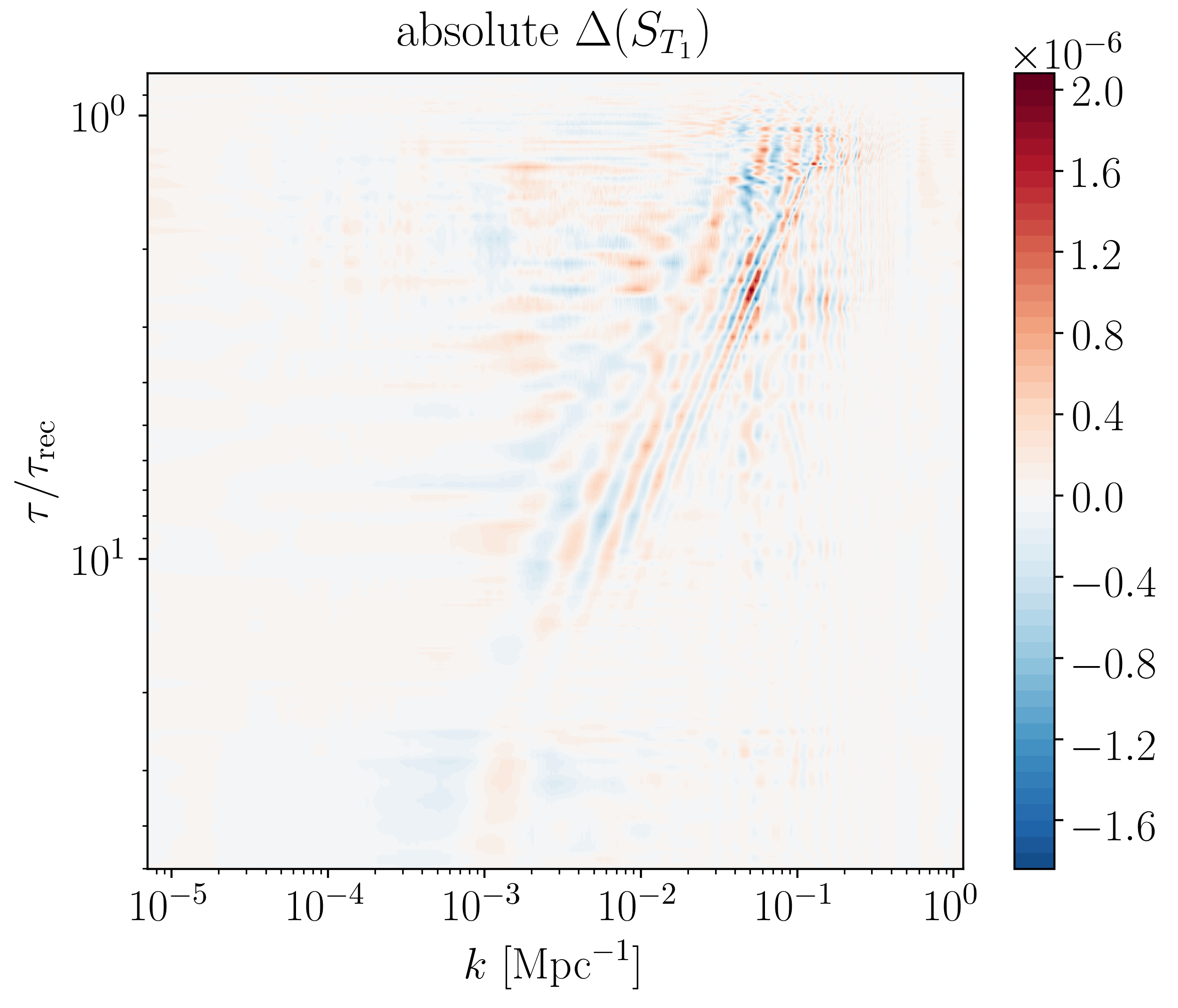}
	\end{subfigure}
	\centering
	\begin{subfigure}[t]{.61\textwidth}
		\centering
		\includegraphics[width=\linewidth]{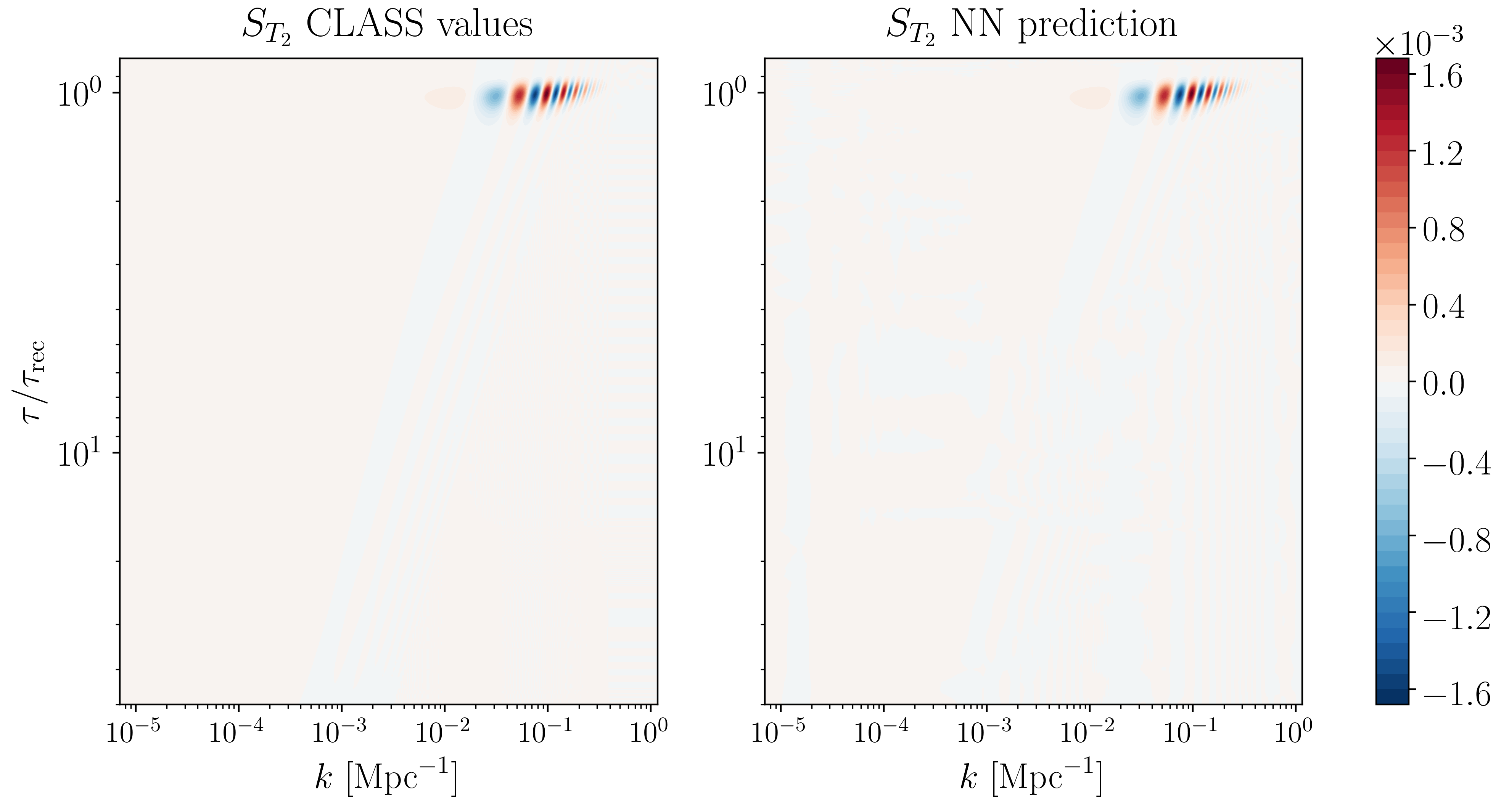}
	\end{subfigure}%
	\begin{subfigure}[t]{.39\textwidth}
		\centering
		\includegraphics[width=\linewidth]{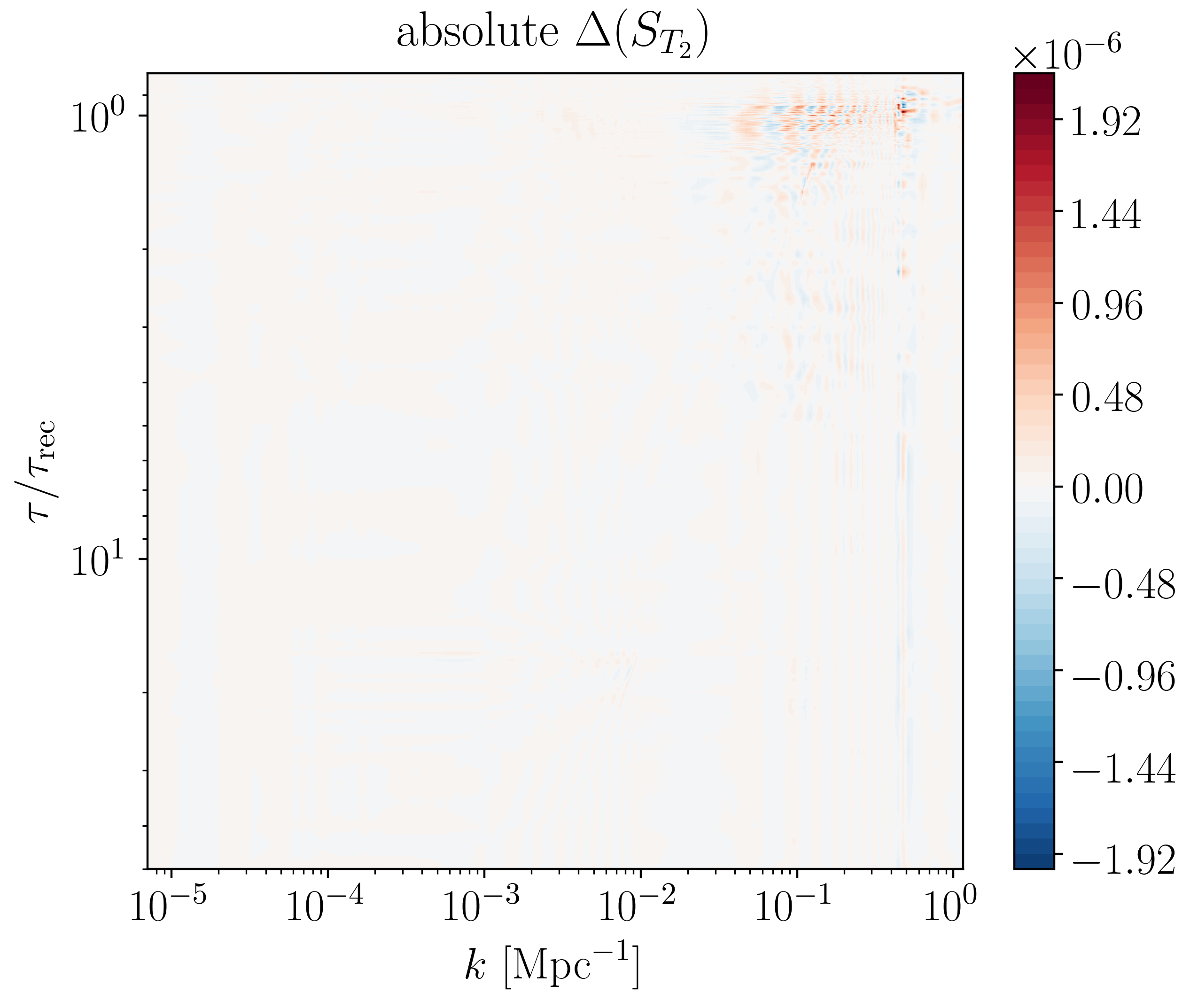}
	\end{subfigure}
	\caption{{\it (Left panels)} Source functions $S_{T_0}(k,\tau)$, $S_{T_1}(k,\tau)$ and $S_{T_2}(k,\tau)$ predicted by \class-full and \class-net for a randomly chosen $\Lambda$CDM model. {\it (Right panels)} Absolute difference between these predictions. Note that the network predicts a rescaled version of the source functions: here the rescaling has been undone.}
	\label{fig:ST0ST1ST2}
\end{figure}
For all subsequent comparisons between \class-full and \class-net predictions, we need a random set of $\Lambda$CDM models. We already used the hyperparameter optimization set of 100 cosmologies to select the best performing network. For the purpose of this section and of \fullsecref{sub:results_power_spectra}, we have generated the \emph{independent} test set of 100 cosmologies, again using Latin Hypercube Sampling. 
\dnew
In the top panels of figure \ref{fig:ST0ST1ST2} we present a comparison of $S_{T_0}$ calculated by \class-full and \class-net using a model picked randomly from this set. The network is able to capture the essential features of the source function and achieves an error that is three orders of magnitude smaller than the peak value. This source function is the most crucial one for the calculation of the temperature anisotropy spectrum.
\dnew
The middle panels of figure \ref{fig:ST0ST1ST2} show the results for $S_{T_1}$. The relative error is slightly bigger in that case because we deliberately chose to reduce the network size and training time for $S_{T_1}$\,, knowing it has an extremely small impact on the final CMB temperature spectrum.
\dnew
\begin{figure}[t]
	\centering{}
	\begin{subfigure}[t]{.61\textwidth}
		\centering
		\includegraphics[width=\linewidth]{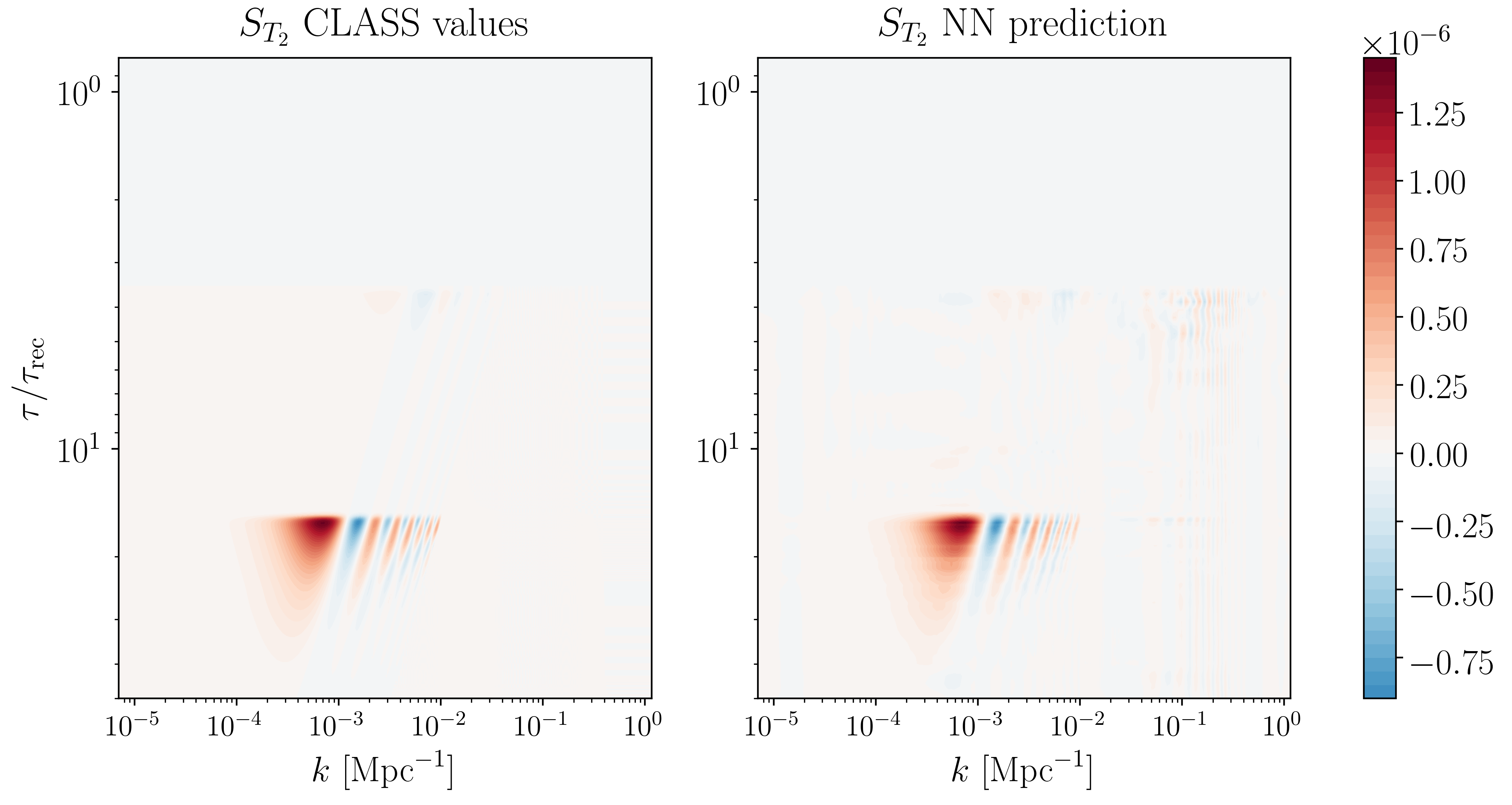}
	\end{subfigure}%
	\begin{subfigure}[t]{.39\textwidth}
		\centering
		\includegraphics[width=\linewidth]{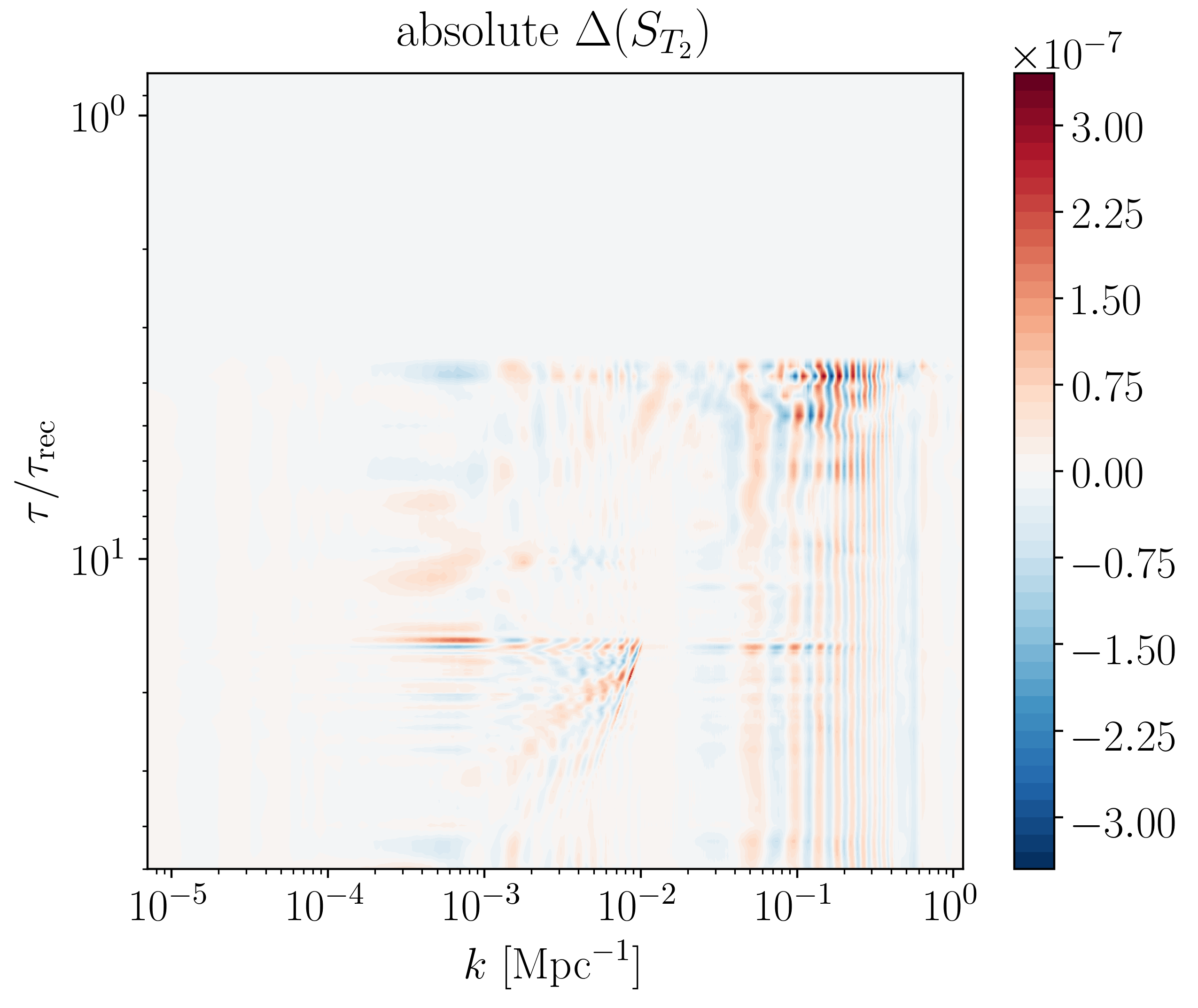}
	\end{subfigure}
	\caption{{\it (Left panels)} Source functions $S_{T_2}(k,\tau)$ predicted by \class-full and \class-net for a randomly chosen $\Lambda$CDM model. Here the structure around the time of recombination has been cut in order to enhance the  structure around the time of reionization, which is three orders of magnitude smaller. {\it (Right panel)} Absolute difference between these predictions.}
	\label{fig:ST2_early_tau_reio}
\end{figure}
\begin{figure}[t]
	\centering
	\begin{subfigure}[t]{.61\textwidth}
		\centering
		\includegraphics[width=\linewidth]{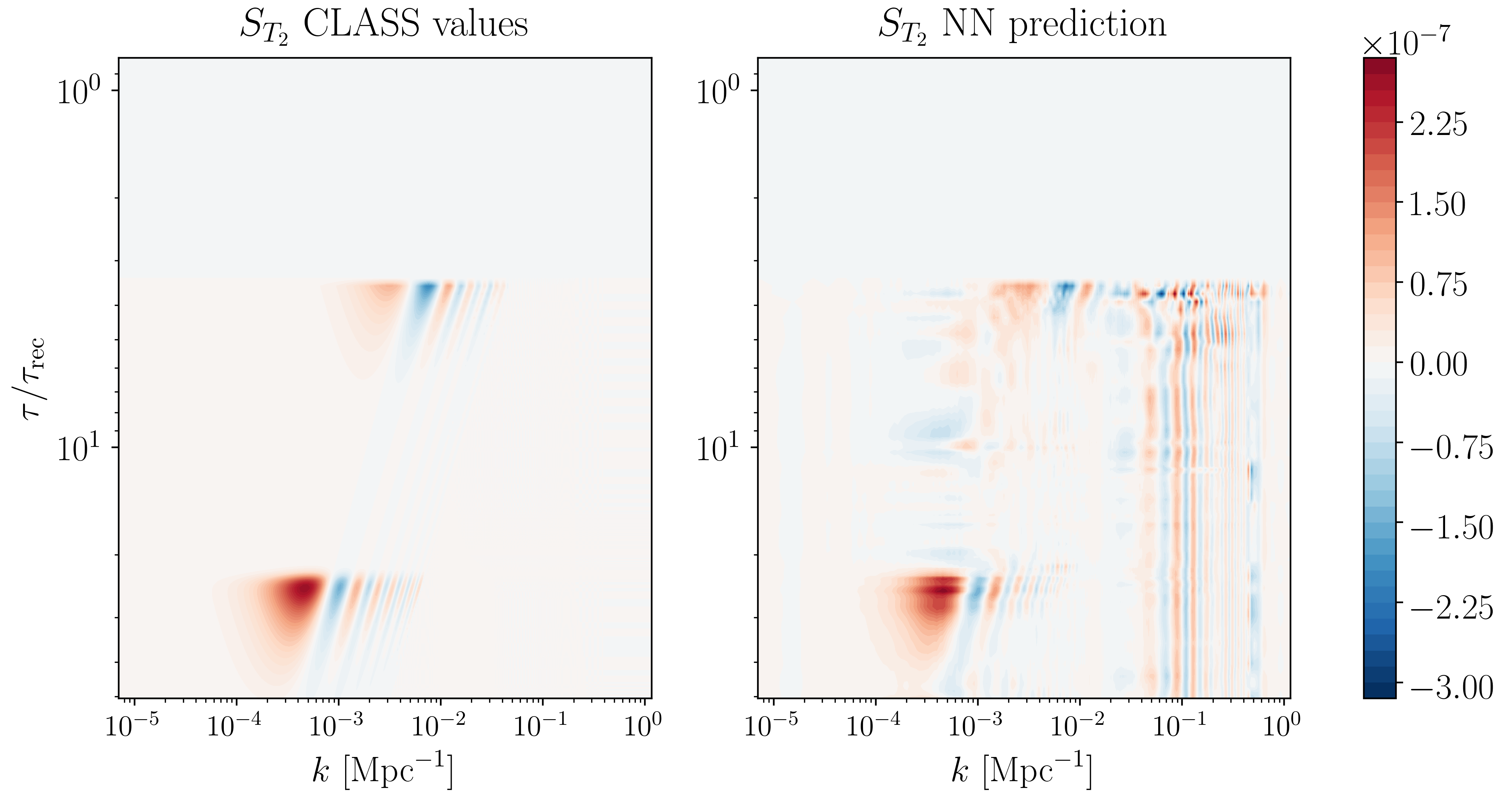}
	\end{subfigure}%
	\begin{subfigure}[t]{.39\textwidth}
		\centering
		\includegraphics[width=\linewidth]{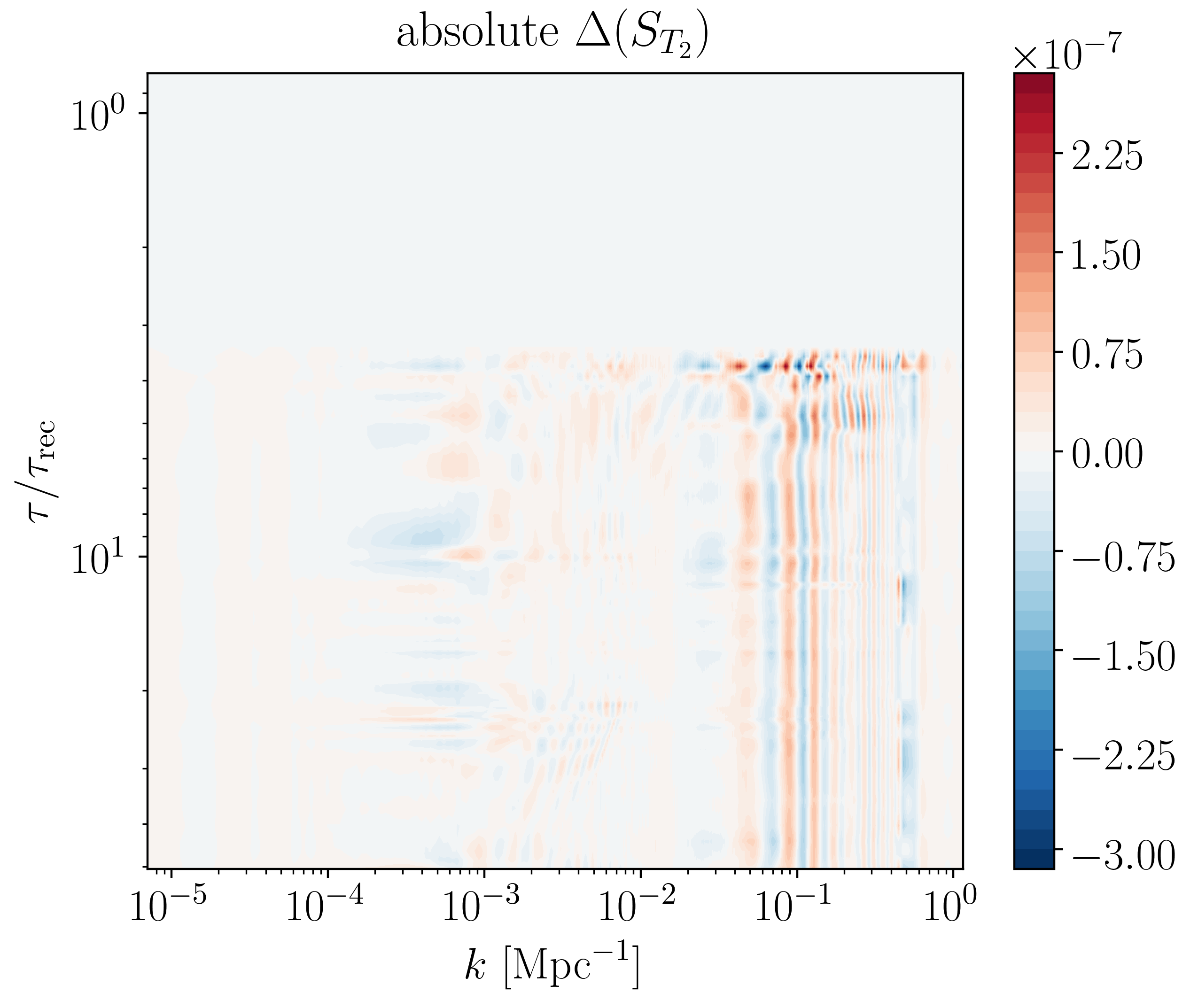}
	\end{subfigure}
	\caption{Same as figure \ref{fig:ST2_early_tau_reio} for a model with very late reionization. In that case the signal is even smaller and the accuracy of the network degrades significantly.}
	\label{fig:ST2_late_tau_reio}
\end{figure}
The source function $S_{T_2}$\,, which has a negligible weight in the temperature spectrum but uniquely determines the polarization spectrum, is predicted with the same accuracy as $S_{T_0}$ (see the lower panels in figure \ref{fig:ST0ST1ST2}). To achieve this result, it was necessary to increase both the training time (epochs) and the density of $\tau$ sampling around $\tau_{\mathrm{reio}}$\,, as touched upon in section \ref{sub:training_procedure}. With such settings the network can accurately capture the non-trivial patterns around the time of reionization, more clearly seen in Figure~\ref{fig:ST2_early_tau_reio}. These patterns determine the shape of the spectra $C_\ell^{EE}$ and $C_\ell^{TE}$ on large angular scales, including the reionization peak.
An additional complication arises from the fact that the signal at $\tau_{\mathrm{reio}}$ becomes even smaller for models with a late time of reionization. Although there is a noticeable drop in quality (see figure \ref{fig:ST2_late_tau_reio}) in these cases, the precision of the polarization anisotropy spectrum does not suffer substantially from this drop, as will be further discussed in \fullsecref{sub:results_power_spectra} and is illustrated in figure \ref{fig:clEE}.

\subsection{Accuracy of the power spectra}
\label{sub:results_power_spectra}

\begin{figure}[t]
	\centering
	\includegraphics[width=\linewidth]{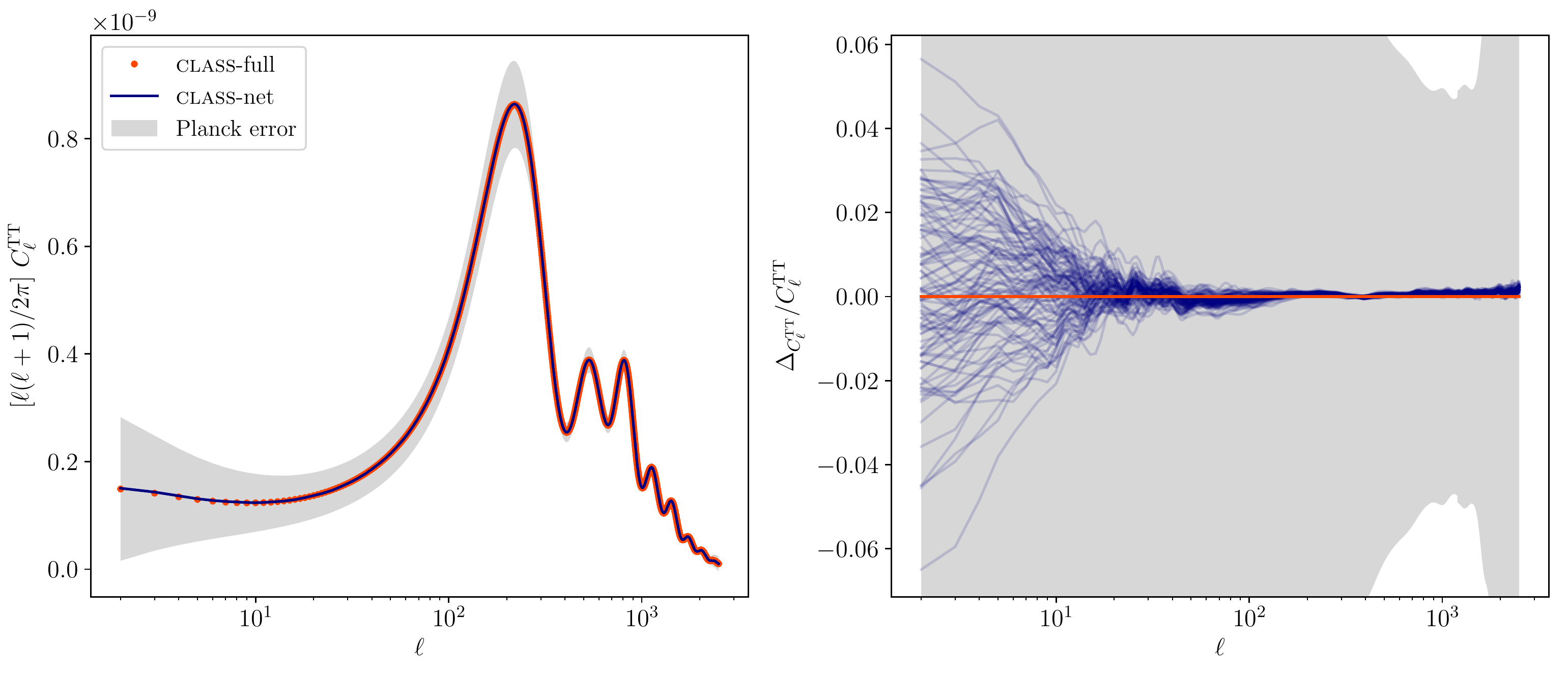}
	\caption{Temperature anisotropy spectrum. The label \class-full represents source functions calculated without the networks, whereas the label \class-net refers to NN-augmented calculations. The figure on the right shows the relative error. The gray band is the experimental error, consisting of cosmic variance for low $\ell$ (large angles) and instrumental noise for high $\ell$ (small angles).}
	\label{fig:clTT}
\end{figure}
\begin{figure}[t]
	\centering
	\includegraphics[width=\linewidth]{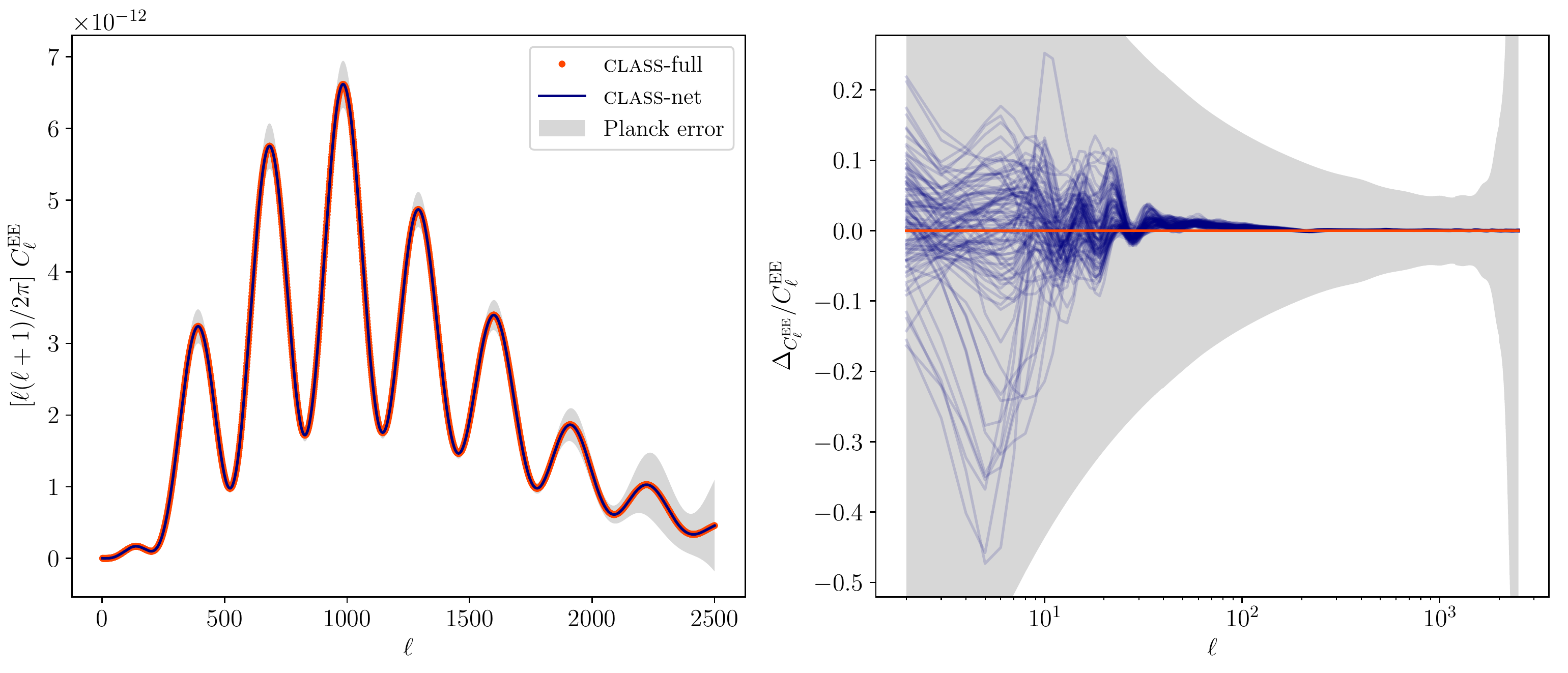}
	\caption{Polarization anisotropy spectrum. The naming conventions of figure \ref{fig:clTT} are adopted. Note that the plot on the left is using a linear $\ell$-axis, while the one on the right uses a logarithmic one.}
	\label{fig:clEE}
\end{figure}
On the left hand side of figure \ref{fig:clTT} we show a direct comparison between the temperature spectrum predicted by \class-full and \class-net for a cosmological model randomly picked from the test set.
The right hand side shows the relative error for the 100 cosmologies of our test set. While there are some apparent systematic effects, the error is small enough for any practical purposes (see next section). The accuracy of \class with default precision settings is of the order of 0.3\% for $C_\ell^{TT}$. The difference between \class-full and \class-net is smaller than 0.3\% for most of the multipole range, but increases on large angular scales and reaches $\sim$\,6\% for $\ell=2$.  This is not worrisome since cosmic variance is very large on these scales. To illustrate this we compare the network error to the (unbinned) Planck error, which we define as the sum of experimental noise and cosmic variance.  The ratio of the two errors is roughly independent of $\ell$, suggesting that the networks have been designed and trained with an efficient weighting between different source functions and different regions in $(k, \tau)$ space. The network error is of the order of $2\%$ of the Planck error.
The same comparison for the $E$-mode polarization spectrum is shown in figure \ref{fig:clEE}. The relative error does reach up~to~$\sim$\,\,40\,\% for very small $\ell < 20$, but stays inside the Planck error boundaries for all $\ell$.
\dnew
The ultimate way to prove that the accuracy of our networks is sufficient given the precision of current data is to perform a full parameter extraction. This is the purpose of the next section. Future generations of CMB experiments may require slightly larger and longer-to-train networks. 

\subsection{Accuracy of parameter extraction from Planck}
\label{sub:results_parameter_extraction}

\begin{figure}[t]
	\centering
	\includegraphics[width=\linewidth]{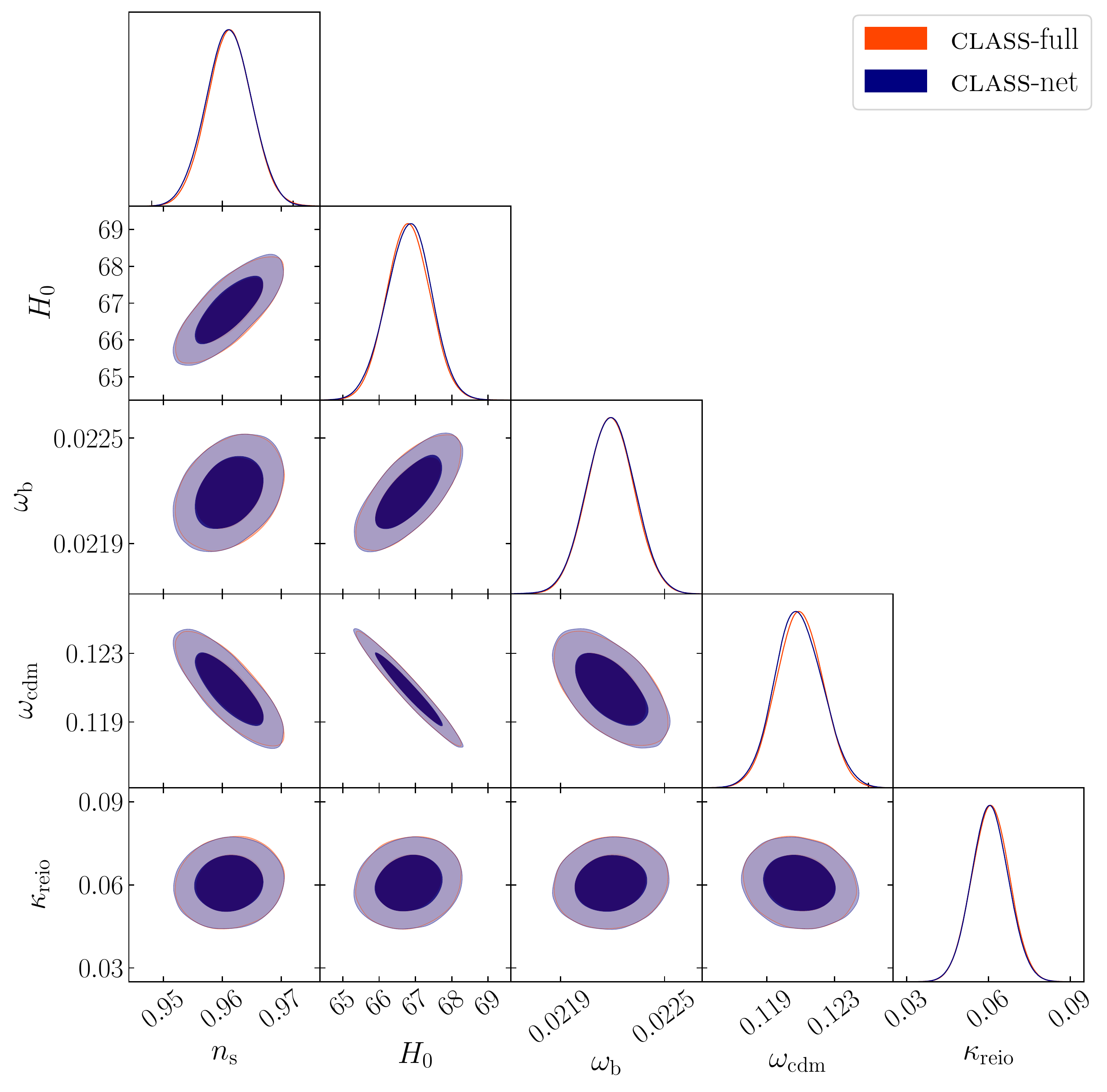}
	\caption{Fit of $\Lambda$CDM parameters to the Planck data, performed either with \class-full or \class-net. We sampled the parameter space with Markov Chain Monte Carlo using the sampler \textsc{cobaya}. 
	}
	\label{fig:cobaya}
\end{figure}

A typical use case for our framework is constraining cosmological parameters using Monte Carlo sampling on cosmological data likelihoods. We present such a computation as a practical benchmark of the accuracy of our NNs, using the code \textsc{cobaya}\footnote{\url{https://github.com/CobayaSampler/cobaya}}. We use a network trained on the $\pm5\sigma$ intervals of the Planck 2018 + BAO posteriors and run an MCMC sample on the Planck 2015 CMB likelihoods (substituting Planck 2015's low-$\ell$ likelihood with a prior on the more restrictive 2018 constraint for $\kappa_{\mathrm{reio}}$, in order to simulate the as-of-today unreleased 2018 likelihoods). Figure~\ref{fig:cobaya} shows the parameter posteriors derived using \class-full and \class-net. The differences, hardly visible by eye, are typically as small as a few percent of the standard deviations. This even suggests that the accuracy of our network could be degraded for speed up without any practical inconvenience.
\dnew
Note that in this work, when running \class-full and \class-net, we always compute the CMB lensing potential by integrating the perturbation equations. Designing networks to also predict this potential (as well as the matter power spectrum) is left for future work. This is in principle a substantially easier task than predicting primary CMB anisotropy spectra, since the transfer functions of matter fluctuation and metric perturbation are considerably smoother than the CMB source functions.

\subsection{Speedup} 
\label{sub:speedup}
The speedup achieved by replacing the perturbations module of \class by our neural networks is measured on the same CPUs on which the training has been performed. In this test we compare the time \class needs for the perturbation module alone, with and without the NN. This is sufficient since the rest of the code remains unchanged. We define the speedup as $S~=~\frac{\text{time(\class-full)}}{\text{time(\class-net)}}$. We run \class with and without the NNs for the one hundred models of our test set in order to get a reliable average. This procedure yields a speedup that can be seen in table \ref{tab:speedup}. In our current implementation the speedup is a factor of 28 when running \class on a single core. Although the speedup decreases with additional cores, it remains significant. With an average execution time of 170\,ms or less, the NN version of the perturbation module is no longer a bottleneck in the EBS. It only contributes at the percent level to the total execution time.
\newpage
We remark here that in this proof-of-concept paper we have focused on size and accuracy of the NN, yielding the almost indistinguishable contours of figure \ref{fig:cobaya}. The present setup leaves plenty of room for optimization of the evaluation speed and its scaling with the number of cores. First, a smaller evaluation time of the NNs is possible if one trades some accuracy for speed. Second, we did not fully optimize the NN evaluations beyond the built-in Keras parallelization. Third, in the speed comparison presented here, the time needed to compute the cosine, sine, and Bessel functions passed as input to the NNs is counted. However we did not fully optimize the numerical method used for this step. We leave all this potential optimization and balancing for future work.
\dnew
Furthermore, the performances presented here refer specifically to a $\mathrm{\Lambda CDM}$ model with one massive neutrino species ($m_{\nu} = 0.06$ eV). For more exotic models where \class has a longer evaluation time, an increased speedup can be expected as the network evaluation time is cosmology-independent. 

\renewcommand{\arraystretch}{1.2}
\begin{table}[t]
	\centering
	\begin{tabular}{cccc}
	\toprule
	Core count & Eval. time \class-full $(s)$ & Eval. time \class-net $(s)$ & Speedup \\ 
	\midrule
	1 & 4.62 & 0.165 & 27.9 \\
	2 & 2.34 & 0.137 & 17.0 \\
	4 & 1.18 & 0.119 & 10.0 \\
	8 & 0.638 & 0.103 & 6.2 \\
	\bottomrule
	\end{tabular}
\caption{Average evaluation times and speedup depending on core-count, calculated across 100 cosmologies out of our test set. Performance analysis was done on Intel Xeon Platinum 8160 processors of the RWTH Aachen Compute Cluster.}	
\label{tab:speedup}
\end{table}
\renewcommand{\arraystretch}{1}


\section{Discussion}
\label{sec:discussion}

This proof-of-concept paper shows that neural networks can speed up EBSs in a decisive way, removing the bottleneck from the integration of perturbation equations, while sticking to relatively shallow networks that can be trained in about half a day on just four cores. With our implementation, the networks only need to be retrained when significantly changing the background, thermal or perturbation evolution by extending the cosmological model or the parameter ranges. However, when adding parameters modeling the primordial spectrum, the non-linear corrections, or some characteristics of the data (e.g. biases or redshift bins), the same network instances can be used without retraining. We refer to \cite{Manrique} for an independent way of emulating EBS predictions for Large Scale Structure, that has been investigated contemporarily to the present work.
\dnew
In this paper we focused on the prediction of the source functions for primary CMB anisotropies. We plan to generalize our approach to the calculation of all transfer and source functions relevant for computing observable spectra, which are those of matter density and velocity perturbations, and metric fluctuations. This extension will be rather straightforward given the smoothness of these source functions. Then the calculation of CMB lensing, cosmic shear and number count $C_\ell$'s (with or without general relativity corrections) will fully benefit from the new method. We will also optimize our numerical implementation of calls to the networks within \class in order to increase performance, especially on multi-core platforms. After these steps, we expect to make the first public release of our NNs together with a version of \class able to switch between the \tquote{full} mode and the \tquote{net} mode. These NNs will not refer specifically to \class and will be ready-to-plug in other EBSs.
\dnew
This new approach has high potential for many longer-term developments and more ambitious goals. 
\dnew
As a first step, the NN based EBS will be released together with user-friendly tools that will allow independent cosmologists to retrain their networks without substantial effort whenever a new parameter (or an extended parameter range) is required. This will induce a shift away from the scenario in which different users run the \tquote{full}-mode EBS for the same model over and over again.
\dnew
However, multiple users will still independently retrain their networks for the same extensions of the baseline model. To avoid this, as a second step we would set up one or several public repositories hosting collections of training sets and trained networks, and scripts to exchange information efficiently between repository and users. When a user would try to explore a new model, a script in their EBS would exchange information with the repository. If the repository contains a version of the networks suited for this new purpose, these networks would be downloaded. If not, the user would still need to generate a new training set and go through a training phase, but their training set -- or directly their newly trained networks -- could be uploaded to the repository, which would then cover more and more models and parameter space volume. The goal of this collaborative approach would be to gradually remove the need for integrating perturbation equations on a global scale, leading to very important savings of computing time and underlying costs.
\dnew
Finally, the neural network approach could be efficiently integrated at the level of MCMC parameter extraction codes. These codes could be the appropriate tools to generate training sets on-the-fly, or to dialogue with the public network repository.
\dnew
We believe that these various futuristic directions are worth exploring and may boost the efficiency of parameter extraction at the global level of the cosmology community. We call this project CosmicNet, where \tquote{Net} refers both to \tquote{neural network} and to a \tquote{collaborative network of cosmologists} mutualizing efforts and optimizing resources in a semi-automatic way thanks to machine learning. 

\vspace*{-0.25\baselineskip}
\section*{Acknowledgements}
We thank Elena Sellentin for open and friendly exchanges about our respective projects. 
NS acknowledges support from the DFG grant LE 3742/4-1. Simulations were performed with computing resources granted by RWTH Aachen University under project thes0551.
\vspace*{-0.25\baselineskip}

\bibliography{references}{}
\bibliographystyle{JHEP}

\end{document}